\documentclass[10pt,twocolumn]{autart}

\usepackage{color,amsmath,amstext,amsfonts,amssymb, mathrsfs,mathtools}
\usepackage{graphicx,algorithm,algorithmic}
\usepackage{tikz}
\usepackage{lipsum,adjustbox}
\usepackage{setspace}
\usepackage{array} 
\usepackage{booktabs}  
\usepackage{bbm}
\usepackage{wasysym}
\usepackage{textcomp}
\usepackage{hyperref} 
\usepackage[sort,compress]{cite}
\usepackage{subcaption} 

\newtheorem{lemma}{Lemma}
\newtheorem{assumption}{Assumption}

\newtheorem{theorem}{Theorem}

\renewcommand{\thesubsubsection}{{\it \Alph{subsection}.\arabic{subsubsection}.}}

\newcounter{l1}
\newcounter{l2}
\newcounter{l3}
\newcommand{\bdotlist}{\begin{list}{$\bullet$}{}}
\newcommand{\bboxlist}{\begin{list}{$\Box$}{}}
\newcommand{\bbboxlist}{\begin{list}{\raisebox{.005in}{{\tiny $\blacksquare$ \ \ }}}{}}
\newcommand{\bdashlist}{\begin{list}{$-$}{} }
\newcommand{\blist}{\begin{list}{}{} }
\newcommand{\barablist}{\begin{list}{\arabic{l1}}{\usecounter{l1}}}
\newcommand{\balphlist}{\begin{list}{(\alph{l2})}{\usecounter{l2}}}
\newcommand{\bAlphlist}{\begin{list}{\Alph{l2}.}{\usecounter{l2}}}
\newcommand{\bdiamlist}{\begin{list}{$\diamond$}{}}
\newcommand{\bromalist}{\begin{list}{(\roman{l3})}{\usecounter{l3}}}

\providecommand{\norm}[1]{\lVert#1\rVert}

\newcommand{\beq}{\begin{equation}}
\newcommand{\eeq}{\end{equation}}

\begin{document}

\begin{frontmatter}

\title{Memory Augmented Neural Network \\ Adaptive Controller for Strict Feedback Nonlinear Systems} 

\thanks[footnoteinfo]{This paper was not presented at any IFAC meeting. Corresponding author Deepan Muthirayan.}

\thanks[footnoteinfo]{Supported in part by the National Science Foundation under Grant Number ECCS-1839429.}

\author[Deepan]{Deepan Muthirayan}\ead{dmuthira@uci.edu},  
\author[Pramod]{Pramod P. Khargonekar}\ead{pramod.khargonekar@uci.edu}  

\address[Deepan]{Electrical Engineering and Computer Sciences, University of California, Irvine, CA 92697} 
\address[Pramod]{Electrical Engineering and Computer Sciences, University of California, Irvine, CA 92697}


%

\begin{abstract}
In this paper, we investigate the adaptive nonlinear control problem for strict feedback nonlinear systems, where the functions that determine the dynamics of the system are unknown. We assume that certain upper bounds for the functions $g_i$s of the system are known. The objective is to design an adaptive controller that can adapt to changes, possibly abrupt, in the unknown functions. We propose a novel backstepping memory augmented neural network (MANN) adaptive control method for solving this problem. The key idea is to augment the controller, in the standard backstepping NN adaptive controllers, with external working memory modules. The controller can write information to its working memory, which in this design is the hidden layer output of the NN, and retrieve this information to modify its output, providing it with the capability to leverage recently learned information to improve its speed of learning. We propose a specific design for this external memory interface. We prove that the proposed control design achieves bounded stability for the closed loop system. We provide numerical evidence on some simulation examples to show that the proposed memory augmentation quite significantly improves the speed of learning and also provide evidence for how memory augmentation improves the speed of learning. 
\end{abstract}

\end{frontmatter}

\section{Introduction}
\label{sec:int}

Human cognition according to Neisser, Cognitive Psychology, 1967, can be defined as ``all processes by which the sensory input is transformed, reduced, elaborated, stored, recovered, and used". It includes perception, memory, attention, reasoning, problem solving, and knowledge representation. Machine learning systems have been significantly enhanced in terms of the range of tasks they can learn to perform and the effectiveness of their learning by incorporation of cognitive elements such as memory and attention \cite{graves2014neural, santoro2016meta, parisotto2017neural}. This naturally leads to the question whether such elements can be incorporated in learning control systems to improve their learning capability. In this paper, we take the first step towards addresing this question by considering adaptive control systems and an external working memory, a specific cognitive capacity. The question we would like to address is whether learning can be improved by augmenting neuro-adaptive controllers with cognitive elements such as a working memory?

Adaptive control theory provides tools and techniques for the synthesis of controllers that can adapt to changes in the parameters in the system dynamics. 
The challenge is to design an adaptive controller such that the closed loop system is stable and matches the desired performance even as system parameters  evolve. Both deterministic and stochastic adaptive control approaches have been widely studied over the last five decades and a great deal of progress in adaptive control has been made that has been documented in the scholarly literature. For the deterministic formulations, the reader is referred to the standard text books \cite{aastrom2013adaptive, lavretsky2013robust, bodson1989adaptive, narendra2012stable, eugene2013robust, krstic1995nonlinear} and references therein. 

In this work, we focus on neural network (NN) based direct adaptive nonlinear control. The literature on NN based adaptive nonlinear control is extensive. The reader is referred to some of the standard text books \cite{miller1995neural, lewis1998neural, lewis2003robot} and \cite{narendra1990identification, polycarpou1991identification, chen1992adaptive, lewis1996multilayer, polycarpou1996stable, narendra1997adaptive, ge1999adaptive, kwan2000robust, zhang2000adaptive, johnson2001neural, ge2004adaptiveMIMO, ge2004adaptive, patre2008asymptotic, hayakawa2008neural, chen2010robust, he2015adaptive, ma2018trajectory} for further reading. Our main idea is a {\em novel architectural modification wherein the NNs are augmented with an external memory module.} This idea arose from scientific discoveries in neuroscience and cognitive science. More specifically, we are inspired by the growing knowledge regarding the role of memory systems in human learning. For example, the paper \cite{gershman2017reinforcement} by Gershman et al. shows how complementing memory systems aid human learning. 

In this paper, we focus on control of a certain class of nonlinear systems, namely strict feedback nonlinear systems. There is a rich history of adaptive control for this class of nonlinear systems. Kanellakopoulos, Kokotovic and Morse
\cite{kanellakopoulos1991systematic} pioneered a recursive design procedure known as the adaptive backstepping controller. They showed that the resulting closed loop system is globally stable and achieves asymptotic tracking. Kanellakopoulos et al. 
 \cite{kanellakopoulos1991systematic} extended the backstepping idea to a much broader class of nonlinear systems called pure-feedback systems, and showed the closed loop system to be regionally stable. Krstic, Kanellakopoulos and Kokotovic 
\cite{krstic1995nonlinear} extended the adaptive backstepping technique to parametric strict-feedback systems. Neural network based adaptive backstepping method was proposed for a class of nonlinear systems Polycarpou 
 \cite{polycarpou1993robust}. The authors showed that the closed-loop system is semi-global stable. 
This was extended to the general strict-feedback system case by Ge, Wang \& Lee \cite{ge2000adaptive}. 

In a very recent paper \cite{muthirayan2019memory}, we introduced a memory augmented neural network adaptive controller for model reference adaptive control (MRAC) and robot arm trajectory tracking controller. In the design proposed in \cite{muthirayan2019memory}, an external working memory is augmented to the NN. The controller can read or write to the memory, very similar to the working memory systems in the human brain. The information that is read from the memory is used to modify the output of the NN, thus serving as a complementing memory system to the NN. In \cite{muthirayan2019memory} we proposed a specific design for the working memory and showed that just by leveraging the information in the memory the learner of the controller was able to respond to abrupt changes quicker than a regular NN control in the feedback look. The speed of learning was improved because the {\it information from the memory provided an inductive bias to the learner that updates the NN parameters}. 
 
 In this work, we extend the memory augmented NN idea to the backstepping NN adaptive control design. Our salient contributions in this paper are (i) {\it design of memory augmented NN adaptive backstepping controller for strict feedback systems} (ii) {\it proof of bounded stability and bounded tracking}, (iii) {\it simulation based evidence to support the idea that the proposed memory architecture and algorithm induces superior learning} and (iv) {\it simulation based evidence for the mechanism of learning}.

In section \ref{sec:prel} we introduce the problem setup and the motivation for controller design. In Section \ref{sec:contarch} we discuss the control architecture and in Sub-section \ref{sec:mem-int} we discuss the memory interface design for the working memory. In section \ref{sec:mannalg}, we introduce the backstepping memorry augmented NN (MANN) adaptive control algorithm, which is based on the Lyapunov stability analysis method proposed in \cite{zhang2000adaptive} and provide stability results. Finally in section \ref{sec:disc} we provide simulation results and a detailed discussion substantiating the improved performance obtained by memory augmentation. 

\section{Problem Preliminaries}
\label{sec:prel}

In this section we briefly discuss the problem setup and the main objective of this paper. 

\subsection{Problem Setup} Denote the state by $x \in \mathbb{R}^n$ and each component of the state by $x_i \in \mathbb{R}$. The plant model is a nonlinear strict feedback system given by equations,
\begin{align} 
\dot{x}_1 & = f_1(x_1) + g_1(x_1)x_2 \nonumber\\
\dot{x}_2 & = f_1(x_1, x_2) + g_1(x_1, x_2)x_3 \nonumber\\
& \vdots \nonumber\\
\dot{x}_n & = f_n(x_1, . . , x_n) + g_n(x_1, . . , x_n)u, \ y = x_1, \label{eq:strictfeedsys}
\end{align}
where $f_i(x_1, . . , x_i)$ and $g_i(x_1, . . ., x_i)$ are unknown functions. We make the assumption that  certain upper bounds of the unknown function $g_i$s are known and that the system state is observable. This assumption is specified in detail below.
\begin{assumption}
(i) $\exists$ strictly positive functions $\mathbf{g}_i(.)$ such that,
\beq  \mathbf{g}_i(.) \geq \vert g_i(.) \vert > g_{i,0} > 0 \eeq
where $g_{i,0}$ is a constant and that $\mathbf{g}_i(.)$ are known functions.\\
(ii) The system state is observable
\label{ass:gfunc}
\end{assumption}

\subsection{Objective and Motivation} The control objective is the same as that of the standard neuro-adaptive controller: the system output $y = x_1$ should track the command signal $y_{d}$ in the presence of uncertainties in the system dynamics. The difference is that in our case the unknown function can also undergo {\it abrupt changes}. The adaptive laws designed for a standard neuro-adaptive controller work well for scenarios where the changes to the unknown function are gradual. The question then is what if the changes are abrupt. In such cases the memory of recent experiences could be leveraged to accelerate learning especially when the abrupt changes are not large. The challenge is to design an algorithm for this purpose and demonstrate that learning can be improved over a regular NN adaptive controller.

In this work, we provide a specific design for an external working memory that augments a NN adaptive controller and show that just by leveraging the information in the memory the learner of the controller is able to respond to abrupt changes much quicker than a regular NN controller. We note that the improvement in learning is just a result of how information from recent learning is leveraged, which is something a regular NN adaptive law is less capable of. We note that this demonstrates the value of {\it architectural innovations inspired from human cognition}. 

\section{Control Architecture}
\label{sec:contarch}

In this section, we introduce the control architecture for the proposed MANN controller and the design of the interface for the working memory. 

\subsection{Control Architecture} The architecture proposed in Fig. \ref{fig:cwmem} is an extension of the standard backstepping NN adaptive control architecture \cite{kwan2000robust}. The standard backstepping controller is the controller in Fig. \ref{fig:cwmem} without the working memory. 
Here, each NN approximator in the feedback loop is augmented with a memory similar to the MANN controller that was proposed in our earlier work \cite{muthirayan2019memory}. The controller can read or write to each of the working memory. The information that is read from each working memory is used to modify the output of the respective NNs. The modified output is fed to the auxiliary control inputs $x_{i,d}$ or the control input $u$ as the case maybe. The state of the system is fed to the error evaluator block which computes the error between states $x_i$s and the corresponding auxiliary control inputs $x_{i,d}$s, as shown in Fig. \ref{fig:cwmem}. The output of the error evaluator are the error signals $e_i$s. These error signals are inputs to the control law which computes the auxiliary control signals $x_{i,d}$s and the final control input $u$. The error evaluator's outputs are also fed to the `udpate law' block which updates the parameters of the NNs. This completes the higher level description of the architecture.

\begin{figure}
  %
  \begin{subfigure}{0.5\textwidth}
     \begin{tikzpicture}[scale = 0.6, transform shape]
\draw [draw=blue, fill=blue, fill opacity = 0.1, rounded corners, thick] (-3, -2) rectangle (3, 1);
\draw (0, -0.5) node[align=center] {$\dot{x}_1 = f_1(x_1)+ g_1(x_1)x_2$ \\ $\dot{x}_2 = f_2(x_1, x_2)+ g_2(x_1, x_2)x_3$ \\ \vdots \\ $\dot{x}_n = f_n(x_1, . . , x_n)+ g_n(x_1, . ., x_n)u$};
\draw (0,-2.5) node[align=center] {Plant};
\draw [draw=blue, fill=blue, fill opacity = 0.1, rounded corners, thick] (-7, -1) rectangle (-4, 0);
\draw (-5.5, -0.5) node[align=center] {Control Input $u$};
\draw [->, thick] (-4,-0.5) -- (-3,-0.5); 


\draw [draw=blue, fill=blue, fill opacity = 0.01, rounded corners] (3, 9.5 ) rectangle (5, 1.5);
\draw (4,10) node[align=center] {Error Evaluator};
\draw [->, thick] ( 4,9.5)  -- ( 4,10.5)  --  (-5.125,10.5) -- (-5.125,9.5);

\draw (4, 8.5) node[circle, draw = blue, thick] (a) { $-$}; 
\draw [->,thick] (a) -- (4,7.75) -- (0.5,7.75) -- (0.5,7.5);  
\draw [->,thick] (2.5,7.75) -- (2.5,6.25);
\filldraw (2.5, 6.25) circle (2pt);
\draw (4, 7) node[circle, draw = blue, thick] (b) {$-$}; 
\draw [->,thick] (b) -- (4,6.25) -- (0.5,6.25) -- (0.5,6);  
\draw [->,thick] (2.7,6.25) -- (2.7,4.75);
\filldraw (2.7, 4.75) circle (2pt);
\draw (4, 5.5) node[circle, draw = blue, thick] (c) {$-$};
\draw [-,thick] (c) -- (4,4.75) -- (2.6,4.75);  
\draw [->,thick] (2.5,4.5) -- (2.5,3.25);
\filldraw (2.5, 3.25) circle (2pt);
\draw (4, 4) node[circle, draw = blue, thick] (d) {$-$};
\draw [->,thick] (d) -- (4,3.25) -- (0.5,3.25) -- (0.5,3); 
\draw (2.4, 4.85) node[align=center] {{\bf $\vdots$}};

\draw [draw=blue, fill=blue, fill opacity = 0.01, rounded corners] (-1, 9.5 ) rectangle (2, 1.5);
\draw (0.5,10) node[align=center] {Control Law};

\draw [draw=blue, fill=blue, fill opacity = 0.1, rounded corners, thick] (-0.5, 8 ) rectangle (1.5, 9);
\draw (0.5, 8.5) node[align=center] {$x_{1,d} = y_{d}$};

\draw [draw=blue, fill=blue, fill opacity = 0.1, rounded corners, thick] (-0.5, 7.5 ) rectangle (1.5, 6.5);
\draw (0.5, 7) node[align=center] {$x_{2,d}$};

\draw [draw=blue, fill=blue, fill opacity = 0.1, rounded corners, thick] (-0.5, 6 ) rectangle (1.5, 5);
\draw (0.5, 5.5) node[align=center] {$x_{3,d}$};

\draw [draw=blue, fill=blue, fill opacity = 0.1, rounded corners, thick] (-0.5, 3.5) rectangle (1.5, 4.5);
\draw (0.5, 4) node[align=center] {$x_{n,d}$};

\draw [draw=blue, fill=blue, fill opacity = 0.1, rounded corners, thick] (-0.5, 2) rectangle (1.5, 3);
\draw (0.5, 2.5) node[align=center] {$u$};
\draw [->, thick] (0.5,2) -- (0.5, 1.25) -- (-5.5,1.25) -- (-5.5,0);

\draw [draw=blue, fill=blue, fill opacity = 0.1, rounded corners, thick] (-8.25, 8.5 ) rectangle (-2, 9.5);
\draw (-5.125,9) node[align=center] {Update Laws};
\draw [->, dashed, thick] (-5.125,8.5) -- (-2,1.5);

\draw [->, thick] (3,0.75) -- (6,0.75);
\draw [->, thick] (3,0.25) -- (6,0.25);
\draw [->, thick] (3,-0.25) -- (6,-0.25);
\draw [->, thick] (3,-1.75) -- (6,-1.75);

\draw (6.25,0.75) node[align=center] (e) {$x_1$};
\draw (6.25,0.25) node[align=center] (f) {$x_2$};
\draw (6.25,-0.25) node[align=center] (g) {$x_3$};
\draw (6.25,-1.75) node[align=center] (h) {$x_n$};

\draw [->, thick] (e) -- (6.25,8.5) -- (a);
\draw [->,thick] (5.5, 0.25) -- (5.5,7) -- (b);
\draw [->,thick] (5.25, -0.25) -- (5.25,5.5) -- (c);
\draw [->,thick] (5.25, -1.75) -- (5.25,4) -- (d);
\draw [->,thick] (1.5,8.5) -- (a);
\draw [->,thick] (1.5,7) -- (b);
\draw [->,thick] (1.5,5.5) -- (c);
\draw [->,thick] (1.5,4) -- (d);

\draw [draw=blue, fill=blue, fill opacity = 0.1, rounded corners, thick] (-5, 3 ) rectangle (-2, 2);
\draw (-3.5,2.5) node[align=center] {NN$_n$}; 
\draw [draw=blue, fill=blue, fill opacity = 0.1, rounded corners, thick] (-8.25,3) rectangle (-5.75, 2);
\draw (-7, 2.5) node[align=center] {\small Working Memory$_n$};
\draw [<->, thick] (-5.75,2.5) -- (-5,2.5);
\draw [->,thick] (-2,2.5) -- (-0.5,2.5);

\draw [draw=blue, fill=blue, fill opacity = 0.1, rounded corners, thick] (-5, 3.5 ) rectangle (-2, 4.5);
\draw (-3.5,4) node[align=center] {NN$_{n-1}$}; 
\draw [draw=blue, fill=blue, fill opacity = 0.1, rounded corners, thick] (-8.25,3.5) rectangle (-5.75, 4.5);
\draw (-7,4) node[align=center] {\small Working Memory$_{n-1}$};
\draw [<->, thick] (-5.75,4) -- (-5,4);
\draw [->,thick] (-2,4) -- (-0.5,4);

\draw [draw=blue, fill=blue, fill opacity = 0.1, rounded corners, thick] (-5, 6 ) rectangle (-2, 5);
\draw (-3.5,5.5) node[align=center] {NN$_2$}; 
\draw [draw=blue, fill=blue, fill opacity = 0.1, rounded corners, thick] (-8.25,6) rectangle (-5.75, 5);
\draw (-7, 5.5) node[align=center] {\small Working Memory$_2$};
\draw [<->, thick] (-5.75,5.5) -- (-5,5.5);
\draw [->,thick] (-2,5.5) -- (-0.5,5.5);

\draw [draw=blue, fill=blue, fill opacity = 0.1, rounded corners, thick] (-5, 7.5 ) rectangle (-2, 6.5);
\draw (-3.5,7) node[align=center] {NN$_1$}; 
\draw [draw=blue, fill=blue, fill opacity = 0.1, rounded corners, thick] (-8.25,7.5) rectangle (-5.75, 6.5);
\draw (-7, 7) node[align=center] {\small Working Memory$_1$};
\draw [<->, thick] (-5.75,7) -- (-5,7);
\draw [->,thick] (-2,7) -- (-0.5,7);
\draw (-5.4, 4.85) node[align=center] {{\bf $\vdots$}};

\end{tikzpicture}
    \caption{Backstepping MANN Adaptive Control Architecture}
    \label{fig:cwmem}
  \end{subfigure}
\end{figure}

\subsection{Memory Interface}
\label{sec:mem-int}

 Denote the memory state corresponding to the $i$th working memory by matrix $\mu_{i} \in \mathbb{R}^{n_s \times N}$, where $n_s$ is the number of memory vectors in the memory $i$ and $N$ is the number of hidden laye units. Denote the output of Memory Read of the $i$th working memory by $M_{i,r} \in \mathbb{R}^N$, the modified NN output of the $i$th NN by $u_{i,ad} \in \mathbb{R}$. Denote the input to the $i$th NN by $\tilde{x}_i $; which is a vector and shall be defined later.  Denote the $j$-th column vector of matrix $\mu_i$ by $\mu_{i,j} \in \mathbb{R}^N$. Below, we briefly discuss the three interface operations, i.e., Memory Write, Memory Read and the NN output for the proposed memory interface.

\subsection{\it Memory Write:} \label{sec:memwrite} In this design, the Memory Write equation for the $i$th working memory is given by,
\begin{align}
 \text{Memory Write:} \ & \dot{\mu}_{i,j} = -z_{i,j}\mu_{i,j} +c_wz_{i,j} a_i +z_{i,j} \hat{W}_{i} e_i \ \nonumber\\
 & z = \text{softmax}(\mu^Tq_i) \label{eq:memorywrite} 
\end{align}

Where $a_i$ is the write vector corresponding to interface $i$, $q_i$ is the query vector for the interface $i$ (to be defined later) and $z_i$ is the vector of weights that determines the relevance of the write vector $a_i$ to the memory vector $\mu_{i,j}$. The write vector $a_i$ for this interface is specified by,
\beq a_i = \sigma_i(V_i^T \tilde{x}_i + \hat{b}_{i,v}) \label{eq:writevec} \eeq
That is, the write vector is set to be the current hidden layer value of the NN. 
In the above equation, $c_w$ is a design constant. We choose this constant to be $3/4$. 

The {\it first term} is the forget term which erases the contents of the working memory at the rate determined by the factor $z_{i,j}$, the $j$th element of vector $z_i$. The {\it second term} updates the contents of the working memory $i$ using the write vector $a_i$. The write vector $a_i$ corresponds to the new information that can be used to update the contents of the memory. The {\it third term} plays a complementary role to the first update term especially when the errors are large.

{\it Discussion}: The weight $z_{i,j}$s are determined by a measure of similarity of the write vector (follows from \eqref{eq:queryvec}) and the memory vectors $\mu_{i,j}$s, which in this design is computed as defined in \eqref{eq:memorywrite}. It follows that the memory vector $\mu_{i,j}$ that is most similar to the write vector $a_i$ is considered eligible for the update. This ensures that the update by the newer hidden layer value, which is the write vector, is consistent with the information already stored at a location $\mu_{i,j}$. Note that both the forgetting and updating occurs at the rate $z_{i,j}$. 

We note that the design proposed here provides a {\it general framework for designing Memory Write operation}. Just that for a different design the write vector has to be defined accordingly. The principle followed for the choosing the write vector is that the type of the information stored in the working memory should match the usage of it. In this design, we choose the write vector to be the hidden layer output because of how the memory contents are used: to modify the hidden layer output of the NN, as described below in Section \ref{sec:nnoutput}. 

\subsection{\it Memory Read:} \label{sec:memread} The Memory Read for the $i$th interface is given by,
\beq \text{Memory Read:} \ M_{i,r} = \mu_i z_i, \ z_i = \text{softmax}(\mu_i^Tq_i) \label{eq:memoryread} \eeq

where $z_i$ is the same vector of weights that determines the similarity of the memory vectors in $\mu_{i}$ to the query $q_i$. Thus, the Memory Read output weighs those memory vectors that are similar to the query the highest in its output. Once again, we note that the design proposed here provides a {\it general framework for designing the Memory Read operation}. In this design, the query vector is specified to be the hidden layer output of NN $i$, i.e.,
\beq q_i = \sigma_i(V_i^T \tilde{x}_i + \hat{b}_{i,v}) \label{eq:queryvec} \eeq 

{\it Discussion}: It follows from what is written in memory $i$ \eqref{eq:memorywrite} and the choice for the query, $q_i$, that the Memory Read operation \eqref{eq:memoryread} retrieves values stored in the memory that are similar to the query $q_i$. 
The specific application will then determine the choice of the query vector. 

Here, the goal is to design controllers that can respond to {\it moderate} abrupt changes quickly. If the abrupt changes are moderate then the learned NNs before the abrupt change, that is an approximation of the unknown functions $f_i$s before the abrupt change, will still be closer to a good approximation after the abrupt change. Hence, the current hidden layer output can be used to retrieve relevant values for aiding the learning just after the abrupt change because it will be closer to a good approximation and so the retrieved values will be closer to a good approximation. Thus, for this design, the query is set to be the current hidden layer output. 

\subsection{\it NN Output:} \label{sec:nnoutput} The learning system (NN) {\it modifies its output} using the information $M_r$ retrieved from the memory. For this memory interface, the NN output is modified by adding the output of the Memory Read to the output of the hidden layer as given below.
\beq \text{NN Output:} \ u_{ad} = - \hat{W}^T\left(\sigma(\hat{V}^T\tilde{x} + \hat{b}_v) + M_r\right) - \hat{b}_w \label{eq:nnoutput} \eeq

We note that this modification determined the specification of the write vector for this design. The modification draws information from an external source, which in this case is the working memory, and so can influence the learning of the NN after an abrupt change. We postulate that the modification as proposed in this paper improves the speed of learning by providing an {\it inductive bias} to the learner, inducing the learner to find a good approximation in quick time. 
This is plausible becuase, as described earlier the Memory Read is designed so as to retrieve values that are relevant. 
In the discussion section we provide empirical evidence that the learning is in fact accelerated through the induced learning mechanism. 

\section{Backstepping MANN Adaptive Control Algorithm and Stability}
\label{sec:mannalg}

In this section, we discuss the derivation of the backstepping MANN control algorithm and provide proof for bounded stability of the closed loop system. First, we discuss the design of the backstepping algorithm for the first order system followed by the design of the algorithm for the more general $n$th order system.  

\subsection{Backstepping Control Algorithm for First Order System}
\label{sec:firstordersys}
In this section, we derive the backstepping MANN control algorithm for the following first order system,
\beq \dot{x}_1 = f_1(x_1) + g_1(x_1)u_1 \label{eq:sysfirstord}\eeq

Define $e = x_1 - y_{d}$ and $\beta_1(x_1) = \mathbf{g}_1 (x_1)/g_1 (x_1)$. Consider the function, 
\beq L_{e_1} = \int_{0}^{e_1} \alpha \beta_1(\alpha + y_d) d\alpha \label{eq:lyapfirstorder} \eeq

We can rewrite $L_{e_1}$ as,
\beq L_{e_1} = e^2_1\int_{0}^{1} \theta \beta_1(\theta e_1 + y_{\text{cmd}}) d\theta \eeq 

 
%

Consider the following control input $u$,
\begin{align}
& u_1 = u_1^* = \frac{1}{\mathbf{g}_1(x_1)}\left( -K_1 e_1 - h_1(\tilde{x}_1) \right)\nonumber \\
& \text{where} \ h_1(\tilde{x}_1) = \beta_1(x_1)f_1(x_1) - \dot{y}_{d} \int_{0}^{1} \beta_1(\theta e_1 + y_{d}) d\theta  \nonumber\\
& \tilde{x}_1 = [x_1, y_{d}, \dot{y}_{d}] 
 \label{eq:continpfirstorder}
\end{align}

We can show that the closed loop system with the control input as defined in  \eqref{eq:continpfirstorder} asymptotically tracks the command signal. We state this as the following lemma.
\begin{lemma}
The closed loop system specified by the plant model \eqref{eq:sysfirstord} and the control input $u_1^*$ is globally asymptotically stable. 
\label{lem:stabfirstorder-1}
\end{lemma}

We refer the reader to the appendix for the proof. In the definition of control input $u_1$, as in \eqref{eq:continpfirstorder}, we assumed knowledge of the function $h_1(\tilde{x}_1)$, which is actually an unknown in our setting. Hence, we consider the approximation to $u_1^*$ as the control input instead, and is given by,
\beq u_1 = \frac{1}{\mathbf{g}_1(x_1)}\left( -K_1e_1 - \hat{h}_1(\tilde{x}_1)\right) \label{eq:continpfirstorder-app} \eeq
where $\hat{h}_1$ is the NN approximation of $h_1$. For the MANN controller, where the NN output is modified according to \eqref{eq:nnoutput}, the approximation $\hat{h}_1$ is given by,
\beq \hat{h}_1 = \hat{W}^T_1 \left(\sigma\left(\hat{V}^T\tilde{x}_1 + \hat{b}_v\right) + M_{1,r}\right)+ \hat{b}_w \eeq

Consider $\hat{W}$ and $\hat{V}$ to be shorthand notation for the weight matrices that includes $\hat{b}_w$ and $\hat{b}^T_v$ in their final rows respectively. Let, 
\beq x_{1,e} = \left[ \begin{array}{c} \tilde{x}_1 \\ 1 \end{array} \right] \ \text{and} \ \hat{\sigma}_1 = \left[ \begin{array}{c} \sigma\left(\hat{V}^Tx_{1,e}\right) \\ 1 \end{array} \right] \eeq

Then, using this shorthand notation we can write $\hat{h}_1$ as, 
\beq \hat{h}_1(\tilde{x}_1) = \hat{W}^T_1\left(\hat{\sigma}_1 + \left[\begin{array}{c} M_{1,r} \\ 0 \end{array} \right] \right) \eeq

For this modified control law \eqref{eq:continpfirstorder-app}, the control gain $K_1$ is no more a simple constant and is set as,
\begin{align} 
& K_1 = K\left(1 + \int_{0}^1 \theta g_1 (\theta e_1 + y_{d}) d\theta \right) \nonumber \\
& + K \left(\norm{x_{1,e}\hat{W}^T_1\hat{\sigma}^{'}}^2_F + \norm{\hat{\sigma}^{'}\hat{V}^T_1x_{1,e}}^2_2 \right) + k_z \norm{\hat{W}_1}_F\norm{\mu_1}_F  \nonumber
\end{align}

We note the difference between the gain defined above and the gain defined in \cite{zhang2000adaptive}. The former includes an additional term: $k_z \norm{\hat{W}_1}_F\norm{\mu_1}_F$. The update laws for the NN parameters are set equal to the standard two-layer NN update laws used in the neural network adaptive control literature \cite{zhang2000adaptive}, \cite{lewis1996multilayer}. 
\begin{align}
\dot{\hat{W}} & = C_w\left(\hat{\sigma} - \hat{\sigma}^{'} \hat{V}_1^Tx_{1,e}\right)e_1 - \kappa C_w\hat{W}_1\nonumber \\
\dot{\hat{V}} & = C_v x_{1,e} e_1 \hat{W}_1^T \hat{\sigma}^{'}  - \kappa C_v\hat{V}_1 
\label{eq:nnupdatefirstorder}
\end{align}

We would like to emphasize that this is not an obvious choice for the NN update laws. The proof for stability reveals why this choice still works even with the inclusion of an external memory. Later, through simulations we show how the inclusion of an external memory significantly improves the learning performance when the system uncertainty undergoes abrupt changes. Below, we establish that the closed loop system specified by the plant, the control law and the NN update laws specified above is uniformly ultimately bounded. The proof follows from proof of Theorem \ref{thm:stability}.

\begin{theorem}
Assume that the memory contents are initialized within a compact set. Suppose that the command signal $y_d$ and its derivative $\dot{y}_d$ are bounded. Then, there exists a sufficiently large $K$ such that the closed loop system specified by the plant model \eqref{eq:sysfirstord}, the control input \eqref{eq:continpfirstorder}, the NN update laws \eqref{eq:nnupdatefirstorder}, the memory interface operations \eqref{eq:memorywrite}, \eqref{eq:memoryread} and \eqref{eq:nnoutput} is uniformly ultimately bounded. 
\label{thm:stabfirstorder-2}
\end{theorem}

\subsection{Backstepping Control Algorithm for $n$th Order System}
\label{sec:nthordersys}
In this section, we discuss the Backstepping MANN controller for the $n$th order system \eqref{eq:strictfeedsys}. For notational convenience, we define $\mathbf{x}_i = [x_1, x_2, . ., x_i]$. Note that the control input $u$ can no more be used to directly control the state variable $x_1$ to track the command signal $y_d$. The state variabe $x_1$ can only be indirectly controlled through the state variable $x_2$. To this end, we define an auxiliary control signal $x_{2,d}$, that the variable $x_2$ has to track. The auxiliary control signal, $x_{2,d}$, is defined as,
\begin{align} 
& x_{2,d} = \frac{1}{\mathbf{g}_1(x_1)}\left( -K_1e_1 - \hat{h}_1(\tilde{x}_1)\right) \nonumber\\
& \hat{h}_1(\tilde{x}_1) = \hat{W}^T_1\left(\hat{\sigma}_1 + \left[\begin{array}{c} M_{1,r} \\ 0 \end{array} \right] \right), \ \tilde{x}_1 = [x_1, y_d, \dot{y}_d]^T \nonumber\\
& K_1 = K\left(1 + \int_{0}^1 \theta g_1 (\theta e_1 + y_{d}) d\theta \right) + k_z \norm{\hat{W}_1}_F\norm{\mu_1}_F \nonumber \\
& + K \left(\norm{x_{1,e}\hat{W}^T_1\hat{\sigma}^{'}}^2_F + \norm{\hat{\sigma}^{'}\hat{V}^T_1x_{1,e}}^2_2 \right)
\label{eq:x2d}
\end{align}
We reiterate that the novelty in our design is the modification of the NN output by the output of the Memory Read $M_{1,r}$ corresponding to the working memory of NN$_1$. As described earlier, $x_2$ should follow the signal $x_{2,d}$ in order to control $x_1$ as desired. As was the case with $x_1$, $x_2$ can only be controlled through the state variable $x_3$ and not directly through an external control input. To this end, we define an auxiliary control input $x_{3,d}$, that $x_3$ has to track. This auxiliary control input $x_{3,d}$ is given by,
\begin{align} 
& x_{3,d} = \frac{1}{\mathbf{g}_2(\mathbf{x}_2)}\left( -K_2e_2 -\mathbf{g}_1 e_1 -\hat{h}_2(\tilde{x}_2)\right) \nonumber\\
& K_2 = K\left(1 + \int_{0}^1 \theta \mathbf{g}_2 (x_1, \theta e_2 + x_{2,d}) d\theta \right) \nonumber \\
& + K \left(\norm{x_{2,e}\hat{W}^T_2\hat{\sigma}^{'}}^2_F + \norm{\hat{\sigma}^{'}\hat{V}^T_2x_{2,e}}^2_2 \right) + k_z \norm{\hat{W}_2}_F\norm{\mu_2}_F, \nonumber\\
& \hat{h}_2(\tilde{x}_2) = \hat{W}^T_2\left(\hat{\sigma}_2 + \left[\begin{array}{c} M_{2,r} \\ 0 \end{array} \right] \right) , \ \tilde{x}_2 = [\mathbf{x}_2, y_d, \dot{y}_d, \ddot{y}_d, \hat{Z}_1]^T, \nonumber
\end{align}
and $\hat{Z}_1$ is the vector of weights of NN$_1$. As before, here too, the NN output is modified by the output of the Memory Read $M_{2,r}$ correspoding to the working memory of NN$_2$. We want $x_3$ to track $x_{3,d}$ and to do so we define another auxiliary control input $x_{4,d}$. This process repeats till the $n$th step where the final control input $u$ is specified. The auxiliary control $x_{k+1,d}$, where $k+1 \leq n$, is given by,
\begin{align}
& x_{k+1,d}  = \frac{1}{\mathbf{g}_k(\mathbf{x}_k)}\left( -K_ke_k -\mathbf{g}_{k-1} e_{k-1} -\hat{h}_k(\tilde{x}_k)\right) \nonumber\\
& K_k = K\left(1 + \int_{0}^1 \theta \mathbf{g}_k (\mathbf{x}_{k-1}, \theta e_k + x_{k,d}) d\theta \right) \nonumber \\
&  + k_z \norm{\hat{W}_k}_F\norm{\mu_k}_F + K \left(\norm{x_{k,e}\hat{W}^T_k\hat{\sigma}^{'}}^2_F + \norm{\hat{\sigma}^{'}\hat{V}^T_k x_{k,e}}^2_2 \right), \nonumber \\
& \text{where}, \ \hat{h}_k(\tilde{x}_k) = \hat{W}^T_k\left(\hat{\sigma}_k + \left[\begin{array}{c} M_{k,r} \\ 0 \end{array} \right] \right), \nonumber \\
& \text{and} \ \tilde{x}_k = [\mathbf{x}_k, y_d, \dot{y}_d, . . , y^k_d, \hat{Z}_1, . ., \hat{Z}_{k-1}]^T.
\label{eq:xk1d}
\end{align}

The function $h_k{(\tilde{x}_k})$ that $\hat{h}_k$ approximates is given by,
\begin{align} 
h_k & = \beta_k f_k(\mathbf{x}_k) + e_k\dot{\mathbf{x}}_{k-1} \int_{0}^{1} \theta \frac{\partial \beta_k (\mathbf{x}_{k-1}, \theta e_k + x_{k,d})}{\partial x_{k-1}} d\theta \nonumber \\
& - \dot{x}_{k,d}\int_{0}^{1} \beta_k (\mathbf{x}_{k-1}, \theta e_k + x_{k,d}) d\theta
 \end{align}

The definition of $h_k$ follows from the design of the backstepping controller. Later, we shall see in the proof for stability of the closed loop system how this is a natural choice for the definition of the function $h_k$. 

Finally, the variable $x_n$ is directly controlled using the plant's control input $u$ to track $x_{n,d}$. The control input $u$ is defined as,
\beq 
u = \frac{1}{\mathbf{g}_n(\mathbf{x}_n)}\left(-K_ne_n -\mathbf{g}_{n-1} e_{n-1} -\hat{h}_n(\tilde{x}_n)\right) 
\label{eq:u}
\eeq

This completes the definition of the control law. The update law for the weights of each NN is set equal to the same update law discussed for the first order system earlier,
\begin{align}
\dot{\hat{W}}_i & = C_w\left(\hat{\sigma}_i - \hat{\sigma}^{'}_i \hat{V}_i^Tx_{i,e}\right)e_i - \kappa C_w\hat{W}_i\nonumber \\
\dot{\hat{V}}_i & = C_v x_{i,e} e_i \hat{W}_i^T \hat{\sigma}^{'}_i  - \kappa C_v\hat{V}_i 
\label{eq:nnupdate}
\end{align}

Below, we establish the stability of the closed loop system with the control law and NN update laws as defined above. 
\begin{theorem}
Assume that the memory contents are initialized within a compact set. Consider the plant model given by \eqref{eq:strictfeedsys}. Let the control law be given by equations \eqref{eq:x2d}, \eqref{eq:xk1d} and \eqref{eq:u}, the NN update laws by \eqref{eq:nnupdate}, and the memory interface operations by \eqref{eq:memorywrite}, \eqref{eq:memoryread} and \eqref{eq:nnoutput}. Suppose that Assumption \eqref{ass:gfunc} is satisfied, the command signal and its derivatives up to order $n$ are bounded, $k_z = K$ and $\kappa = 1/\sqrt{K}$, then there exists (a sufficiently large) $K$ such that the resulting closed loop system is uniformly ultimately bounded.
\label{thm:stability}
\end{theorem}
We refer the reader to the appendix for the proof. 

\section{Discussion and Simulation Results}
\label{sec:disc}

In this section, we provide a detailed illustration and a discussion on the performance of the MANN controller and provide evidence for the mechanism that accelerates learning by considering examples of strict feedback systems. The controller parameter values used for both NN and MANN controller were set to be identical in all the simulation examples discussed below. What we observe in the simulations is that the MANN controller significantly improves the recovery time of the closed loop system, while the peak deviations remain below the deviation observed for the controller without memory. We attribute this to the ability of the MANN controller to quickly learn the new unknown function after an abrupt change leveraging the information in the working memory. We provide empirical evidence for substantiating this claim.

\subsection{Illustration using a Second Order System}
In this example, we consider the $2$nd order system specified by, $f_1(x_1) = 0.1(-1/2x_1+ x_1^2)$ and $f_2(x_2) = 0.1(-0.5 x_2 + x_2^2)$, $g_1(x_1) = 1+0.1x_1^2$, $g_2(x_2) = 1+0.1x_2^2$. For this example we assume that the known upper bound of the function $g_i$s, $\mathbf{g}_i = g_i$. The number of hidden layer neurons and the number of memory vectors are set as $6$ and $1$ respectively. The control gain is set as $K = 20$. The learning rates of the NN update laws are set as $C_w = C_v = 10,  \kappa = 0, k_z = 0$. It was necessary to set $k_z = 0$ to ensure that the contral gains $K_k$ are identical for both the MANN controller and the regular NN controller so that the true effect or the influence of working memory can be ascertained. 

We consider couple of scenarios to illustrate the performance and to provide the comparison between MANN controller and the regular NN controller. In scenario $1$, the command signal $y_d = 0.1$ and the system undergoes the following sequence of abrupt changes,
\begin{align}
f_i & \rightarrow 20 f_i \ \text{at} \ t = 5, \ f_i \rightarrow 2 f_i \ \text{at} \ t = 10 \nonumber\\
f_i & \rightarrow  1/40 f_i \ \text{at} \ t = 20 
\label{eq:scen1}
 \end{align}
 
\begin{figure}[htp]
\begin{tabular}{ll}
\includegraphics[scale = 0.22]{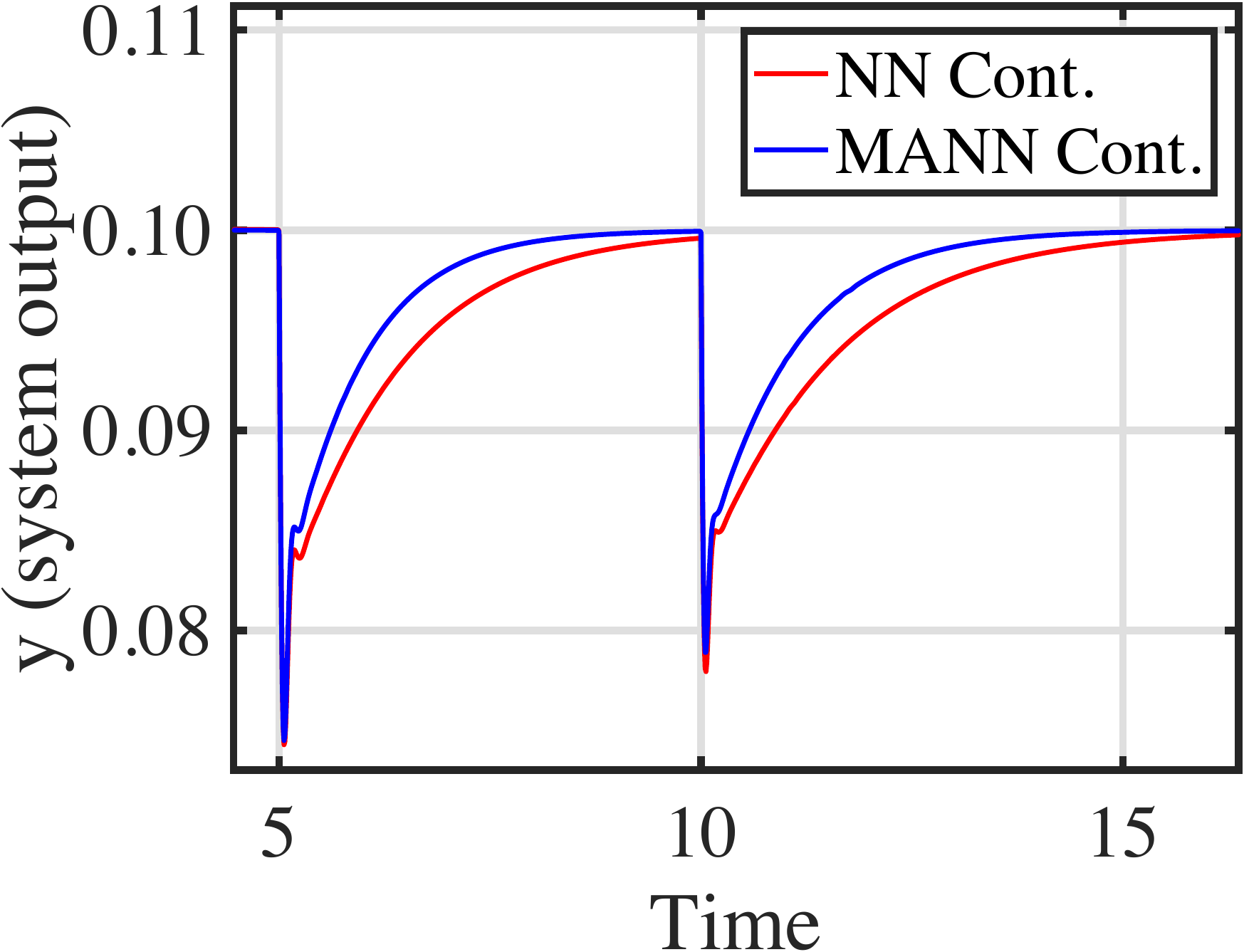} & \includegraphics[scale = 0.22]{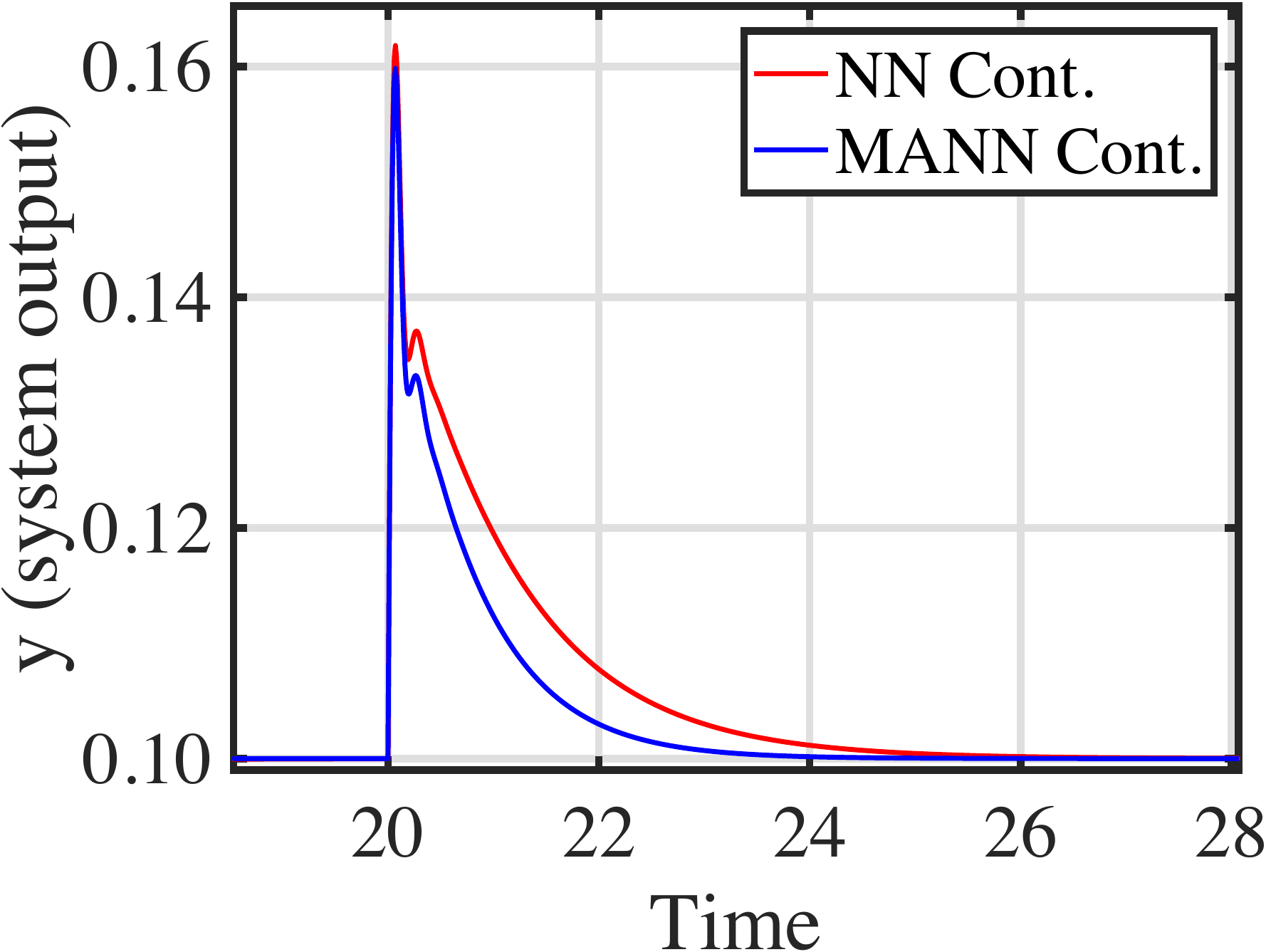}\\
\includegraphics[scale = 0.22]{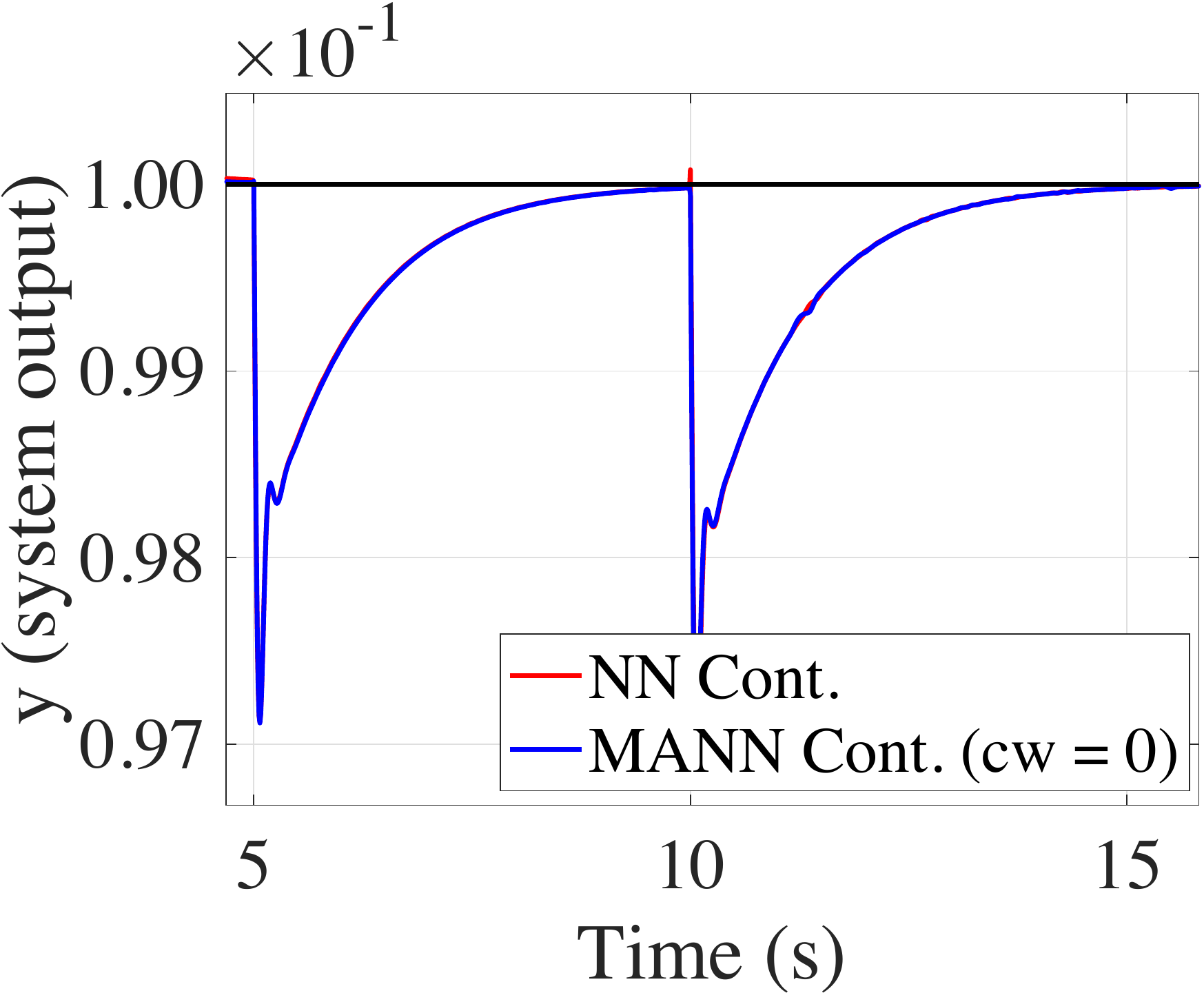} & \includegraphics[scale = 0.22]{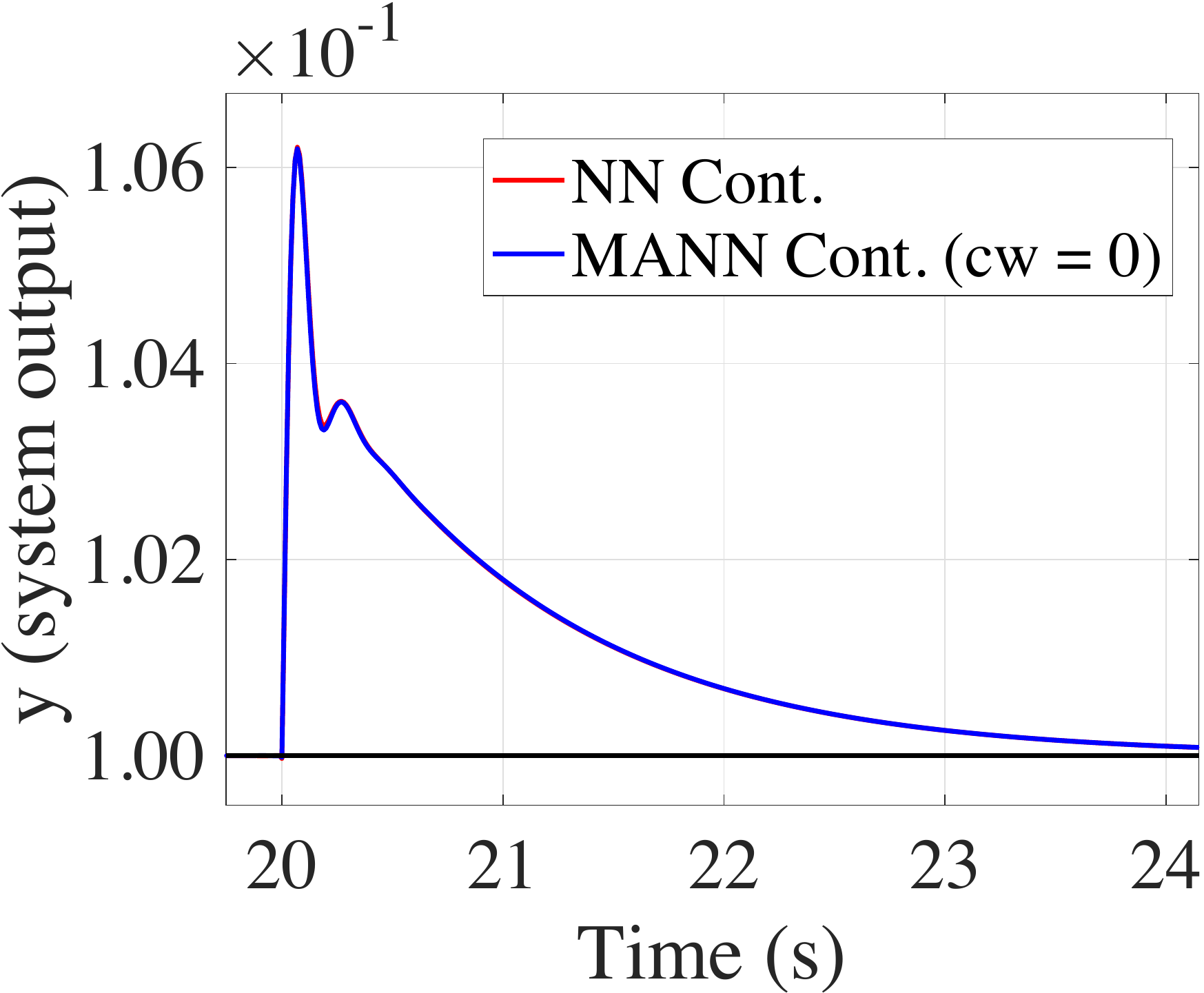}
\end{tabular}
\caption{System response $y$ for example $1$ and scenario $1$. Top left: system response around first and second abrupt changes, top right: system response at the final abrupt change, bottom left: system response around the first two abrupt changes when $c_w = 0$, bottom right: system response around the last abrupt changes when $c_w = 0$}
\label{fig:secordsys-ex1-sc1}
\end{figure}

\begin{figure}[htp]
\begin{tabular}{ll}
 \includegraphics[scale = 0.22]{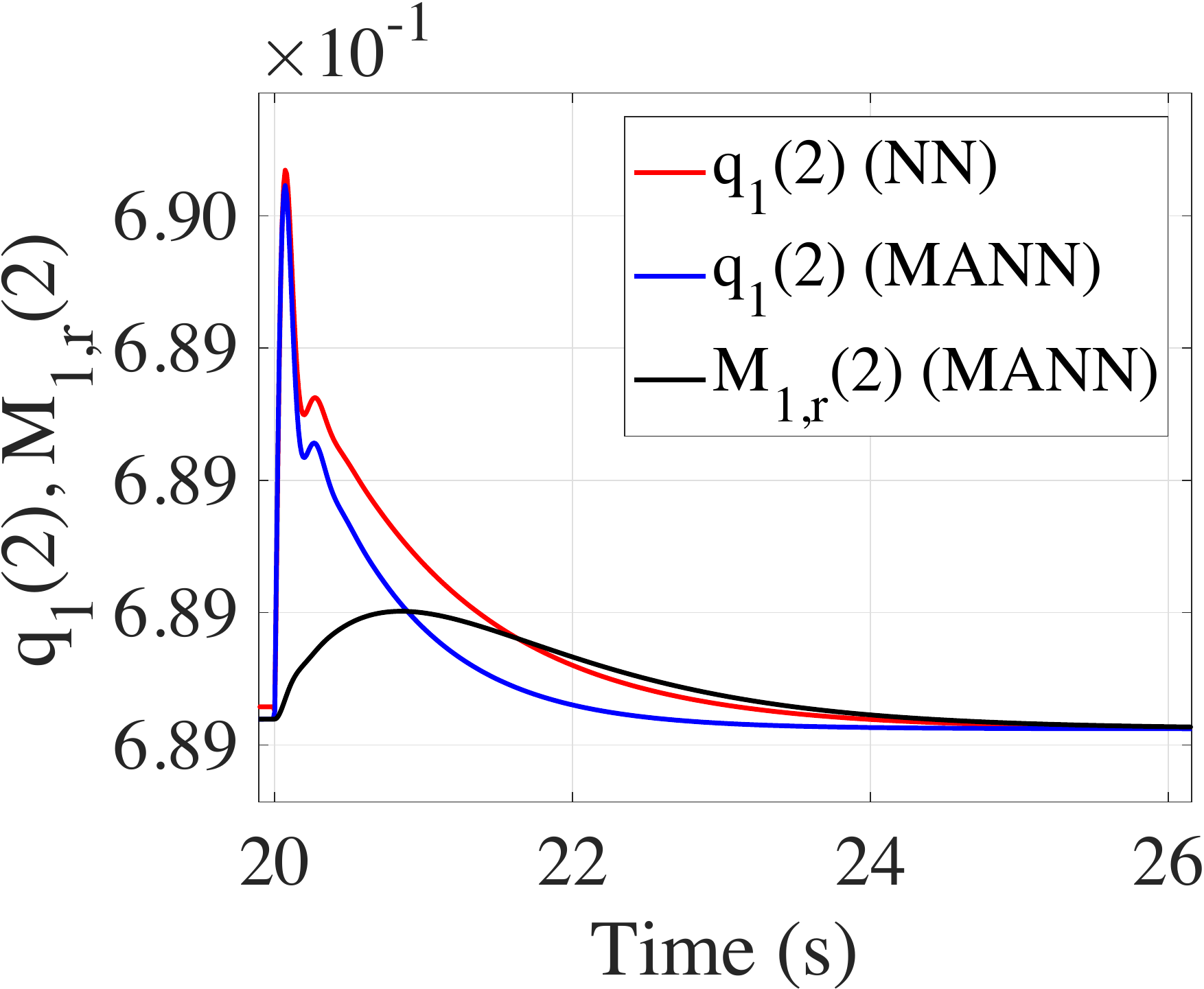} & \includegraphics[scale = 0.22]{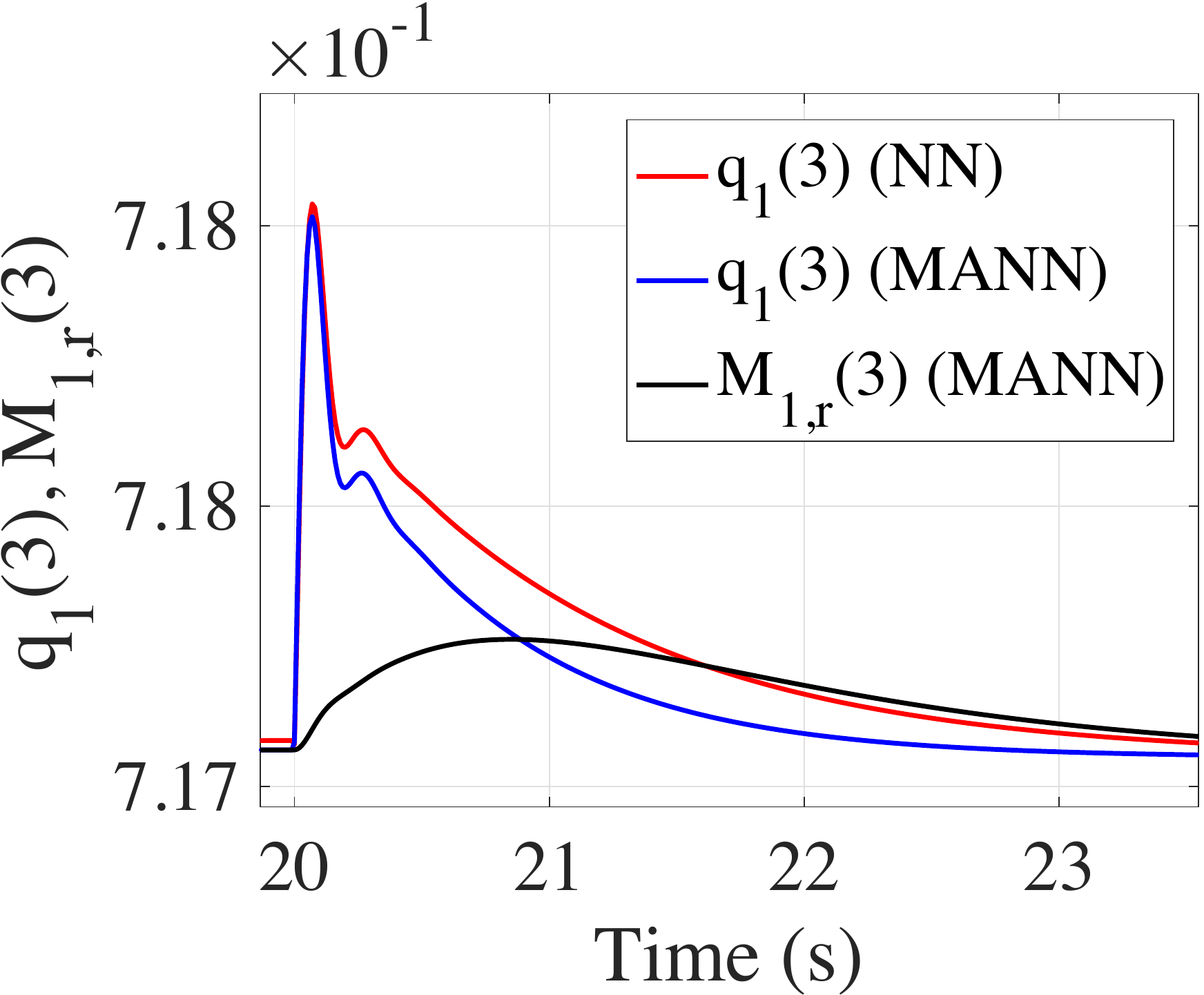} \\
 \includegraphics[scale = 0.22]{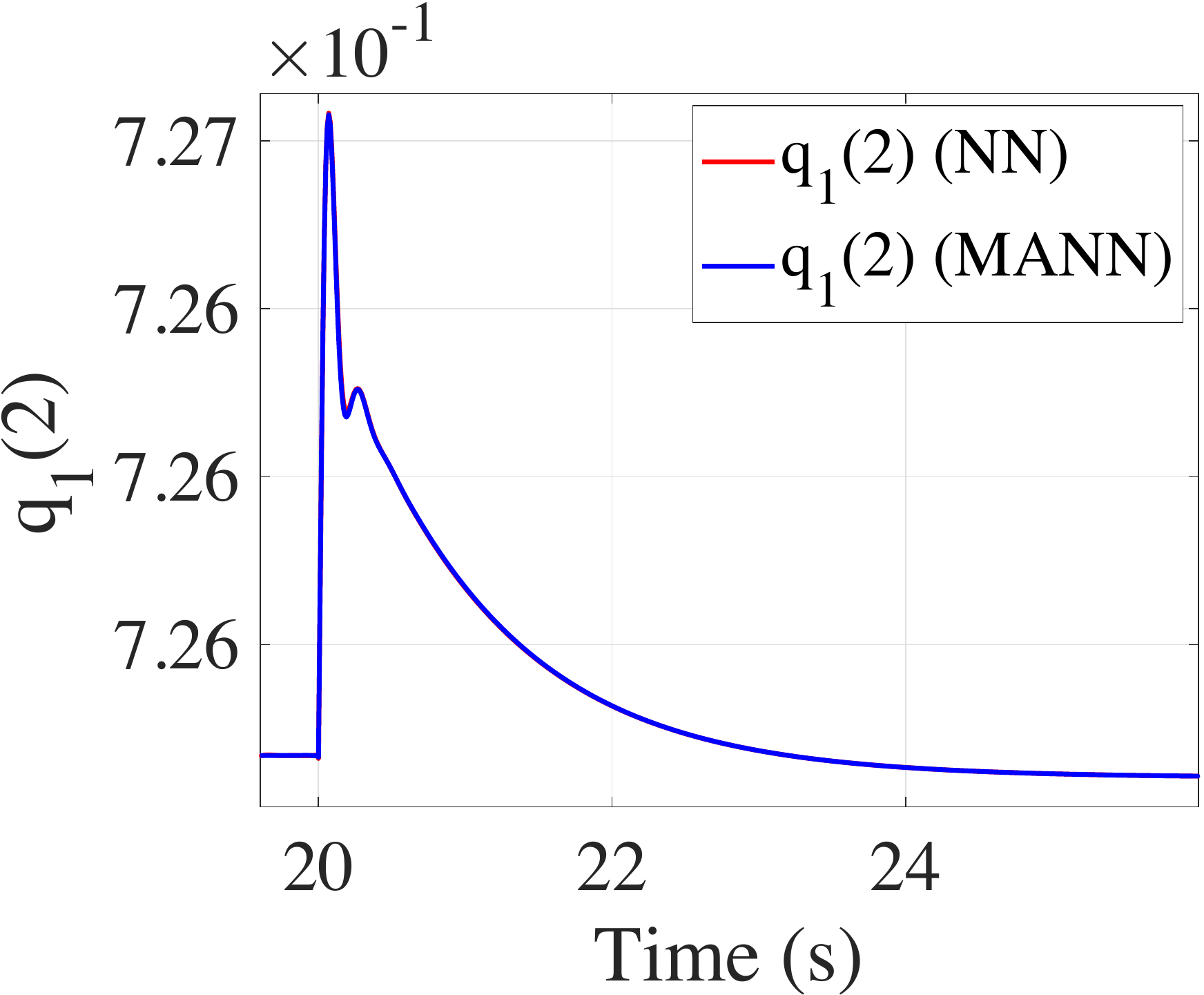} & \includegraphics[scale = 0.22]{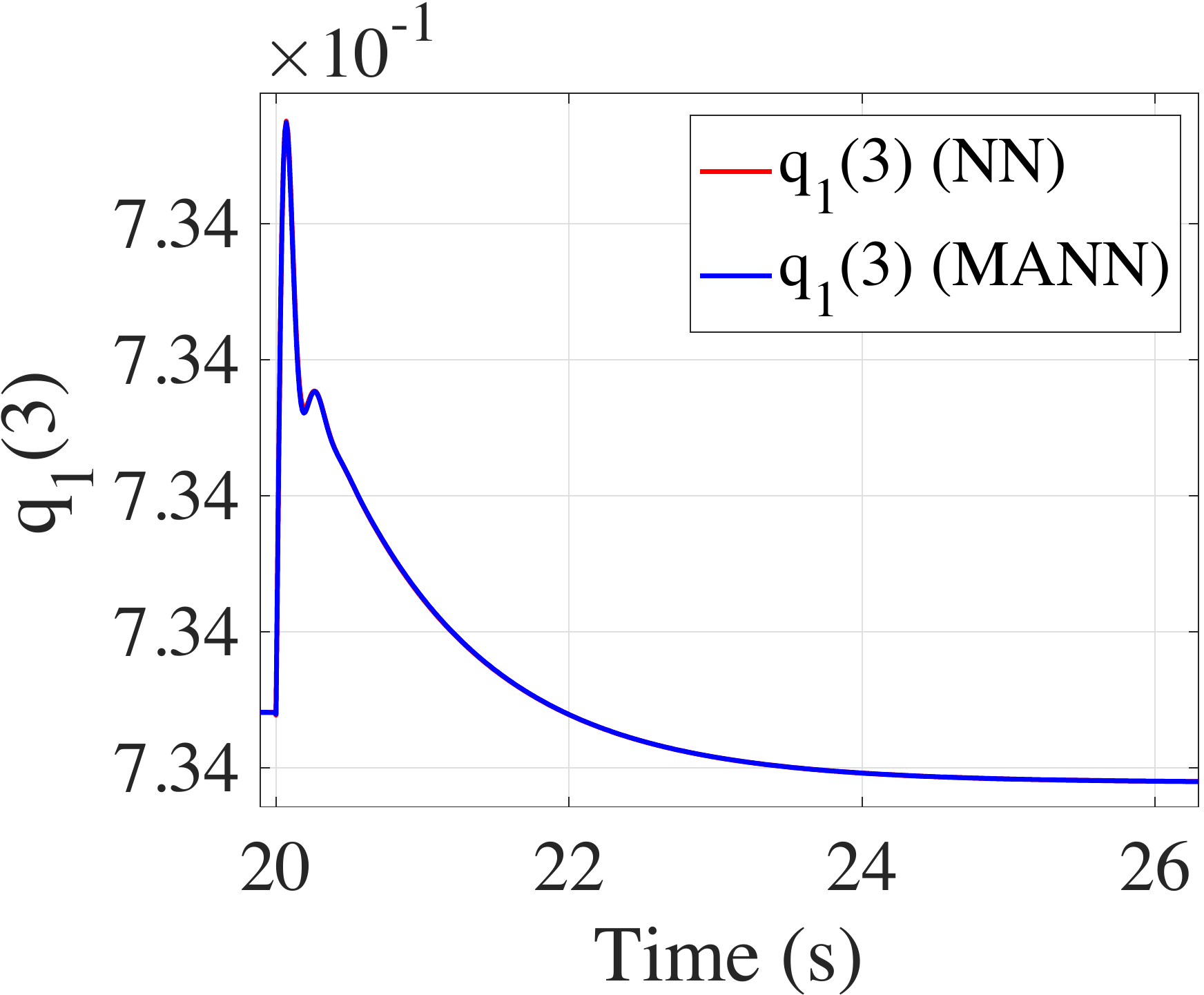}
\end{tabular}
\caption{Plots for example $1$ and scenario $1$ at third abrupt change. Top left: plot of $q_1(2)$ and $1/c_wM_{1,r}(2)$, top right: plot of $q_1(3)$ and $1/c_wM_{1,r}(3)$, bottom left: plot of $q_1(2)$ when $c_w = 0$, bottom right: plot of $q_1(3)$ when $c_w = 0$}
\label{fig:secordsys-ex1-sc1-error-mem}
\end{figure}

The simulation results for this scenario are shown in Fig. \ref{fig:secordsys-ex1-sc1}. The top two plots show the response for the MANN controller for the parameters outlined earlier (call it regular) and the bottom two plots show the response for the MANN controller with $c_w = 0$. In Table \ref{table:settime-sc1-secch} we provide the recovery time for the error to settle within $0.1\%$ error. It is clear that the MANN contoller reduces the recovery time by a significant margin. 
In addition, the plots reveal that the peak deviations do not overshoot the peak deviations corresponding to the controller without memory. 

\subsubsection{Evidence that Improvement in Learning is Induced by the Working Memory}

Setting $c_w = 0$ is equivalent to not updating the working memory with new information (refer \eqref{eq:memorywrite}). From the plots, it is clear that the system with the regular MANN controller in the feedback loop recovers faster after every abrupt change, while the system with the MANN controller with $c_w = 0$ (MANN-w) in the feedback loop does not recover any faster than the regular NN controller without the memory. Noting that the control gains and the learning rates are identical this clearly suggests that learning is improved as a result of augmenting the NN output by the working memory's output whose contents are continuously updated. Below, we provide evidence for how this augmentation improves the speed of learning. 

\subsubsection{Evidence for the Mechanism of Learning}

In Fig. \ref{fig:secordsys-ex1-sc1-error-mem} the plots for the NN hidden layer values for the second and third elements,  $q_1(2)$ , $q_1(3)$, and the corresponding element values of the memory vector scaled by $1/c_w$, i.e. $1/c_wM_{1,r}(2)$ and $1/c_wM_{1,r}(3)$ around the third abrupt change are shown. The plots suggest a plausible explanation for how the response for the MANN controller converges faster. 
{\it First}, we observe that the hidden layer values of the NN for the MANN controller converges faster to the final values when compared to the NN values for the NN controller which does not have an external memory. {\it Second}, the bottom two plots, which are the plots for $q_1(2)$ and $q_1(3)$ when $c_w=0$, clearly indicate that when the memory is not continuously updated, the hidden layer values do not converge any faster than that of the NN controller. The observations are clearly suggesting that {\it when the NN output is modified by the information in working memory that is continuously updated, the learner of the controller is able to leverage this information through the modified NN output to converge to the final NN values in quick time}. 


%
%

\begin{table}[h]
\centering
\caption{Time to settle within $0.1\%$ error}
\begin{tabular}{|c|c|c|}
\hline
Example $1$ (Scenario $1$) & 2nd change & 3rd change\\
\hline
NN cont. (I) & 3.5 & 3.67 \\
\hline
MANN Cont. (II) &  2.28 & 2.43\\
\hline
Reduction (from (I)) & 35 \% & 34\% \\
\hline
\end{tabular}
\label{table:settime-sc1-secch}
\end{table} 

We consider a second scenario, where the abrupt changes are additive in nature. Here the function $f_i$ undergoes the following sequence of abrupt changes:
\begin{align}
& f_i \rightarrow f_i + 0.001 \ \text{at} \ t = 0, \nonumber \\
&  f_i \rightarrow f_i + 0.05 - 0.001 \ \text{at} \ t = 5, \nonumber\\
& f_i \rightarrow f_i + 0.1 - 0.05 \ \text{at} \ t = 10,\nonumber \\
&  f_i \rightarrow f_i + 0.001 - 0.1\ \text{at} \ t = 20 
\label{eq:scen3}
\end{align}

\begin{figure}[htp]
\begin{tabular}{ll}
\includegraphics[scale = 0.22]{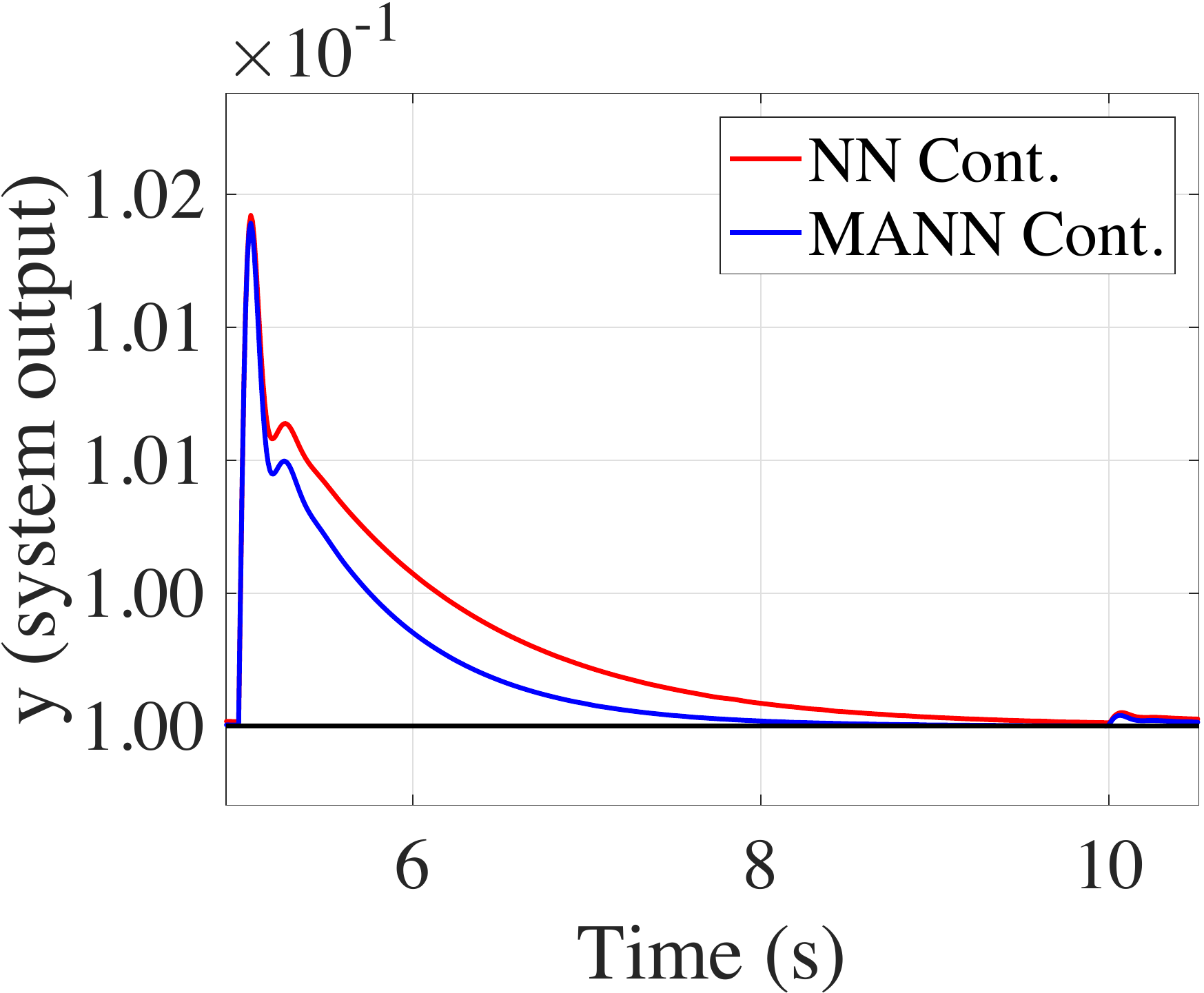} & \includegraphics[scale = 0.22]{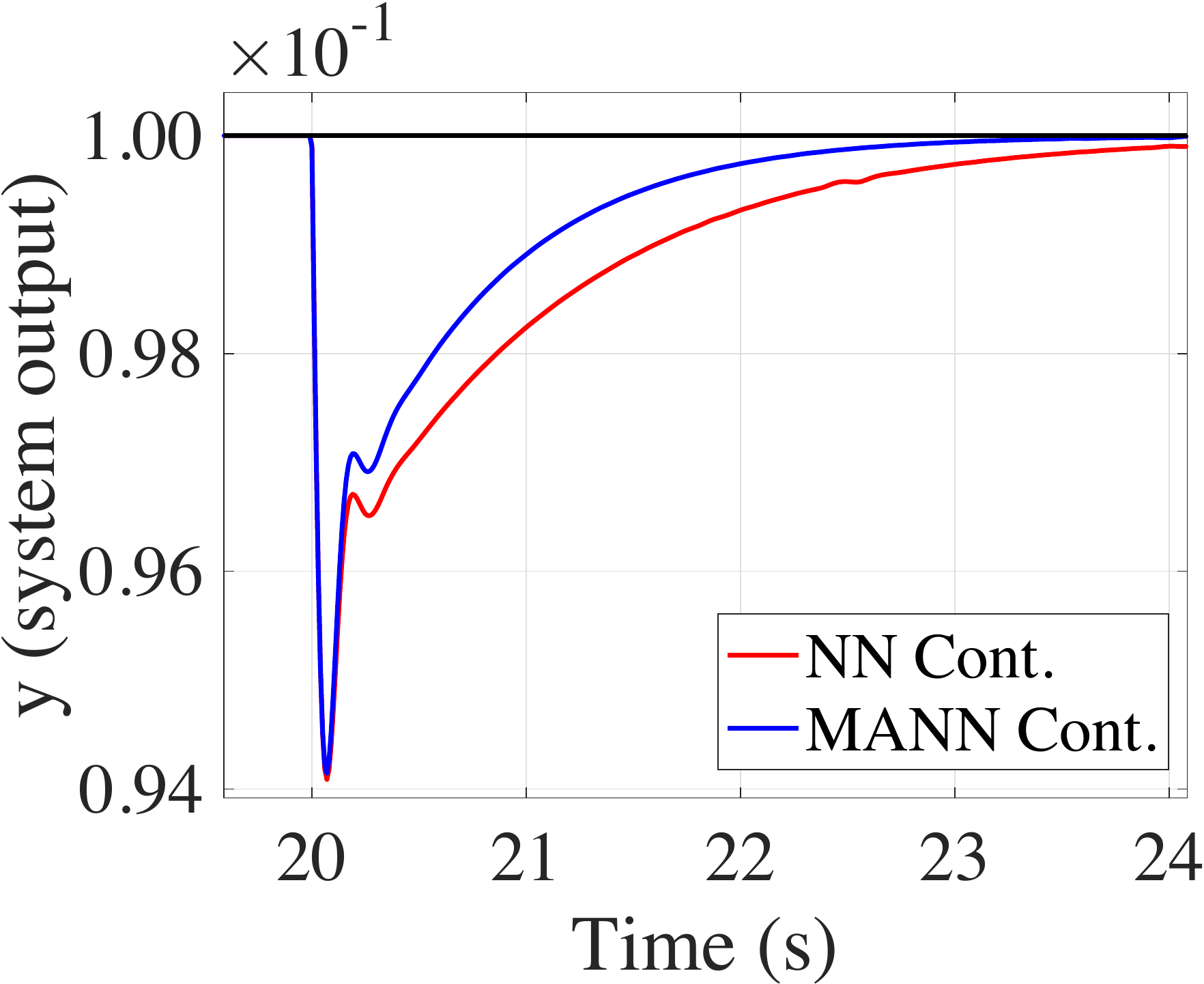}
\end{tabular}
\caption{System response $y$ for example $1$ and scenario $2$. Left: system response around the first two abrupt changes, right: system response around the last abrupt change}
\label{fig:secordsys-ex1-sc3}
\end{figure}

The response of the closed loop system for this scenario and the two controllers are shown in Fig. \ref{fig:secordsys-ex1-sc3}. From the response plots, it follows that the conclusions drawn in the previous scenario apply here as well. Table \ref{table:settime-sc3} lists the values for the time to settle within $0.1\%$ error for both the controllers. It is evident that the MANN controller improves the time to settle by a significant margin for this scenario as well.

\begin{table}[h]
\centering
\caption{Time to settle within $0.1\%$ error}
\begin{tabular}{|c|c|c|}
\hline
Example 1 (Scenario $3$) & 1st change & 3rd change \\
\hline
NN cont. (I) &  2.43 &  3.97 \\
\hline
MANN Cont. (II) & 1.6 & 2.63 \\
\hline
Reduction (from (I)) & 34\% & 34\%\\
\hline
\end{tabular}
\label{table:settime-sc3}
\end{table} 

We also consider a third scenario:
\begin{align}
f_1 & \rightarrow 200 f_1 \ \text{at} \ t = 5, \ f_1 \rightarrow 2 f_1 \ \text{at} \ t = 10 \nonumber\\
f_1 & \rightarrow  1/400 f_1 \ \text{at} \ t = 20 
\label{eq:scen4}
 \end{align}

\begin{figure}[htp]
\begin{tabular}{ll}
\includegraphics[scale = 0.22]{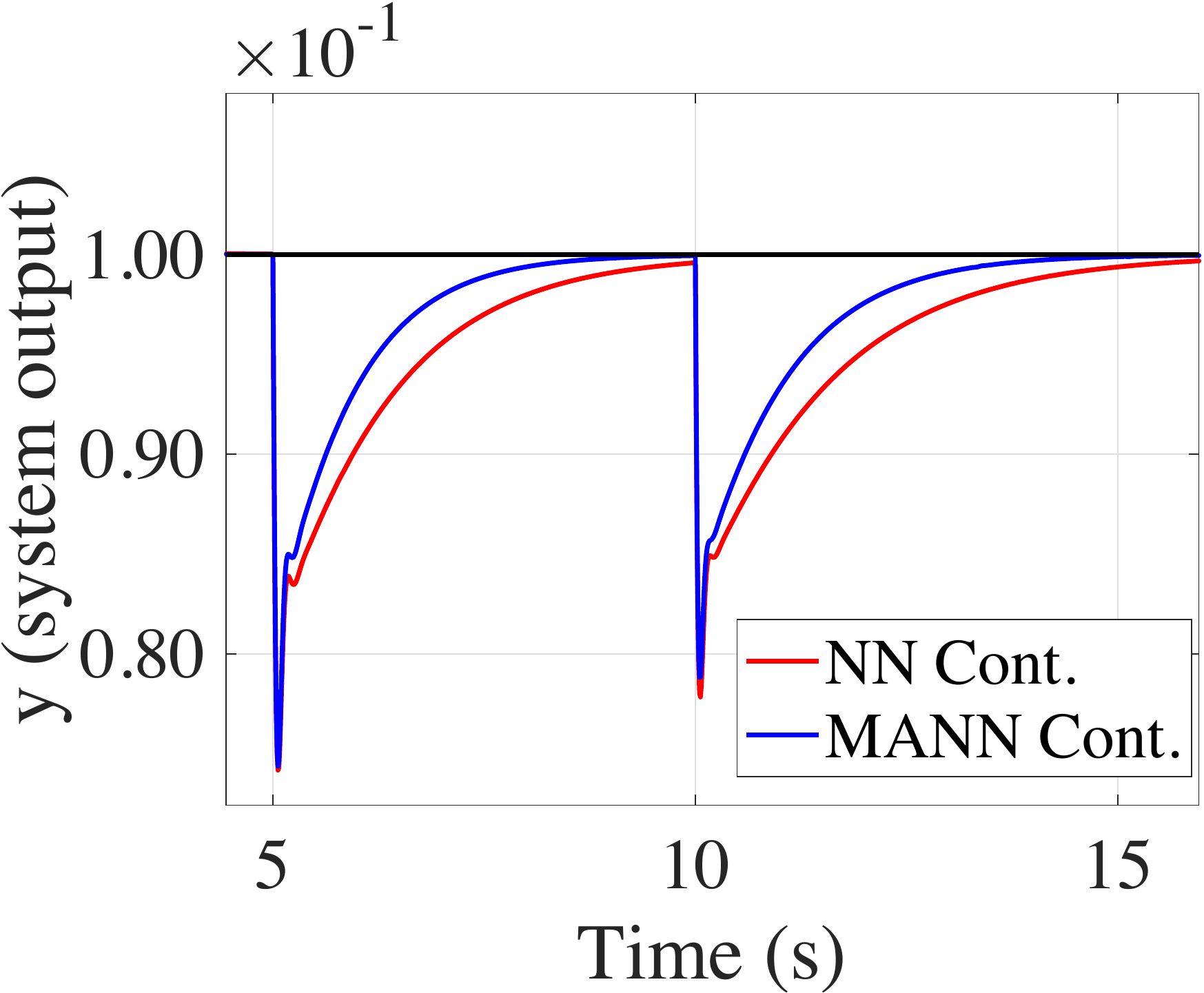} & \includegraphics[scale = 0.22]{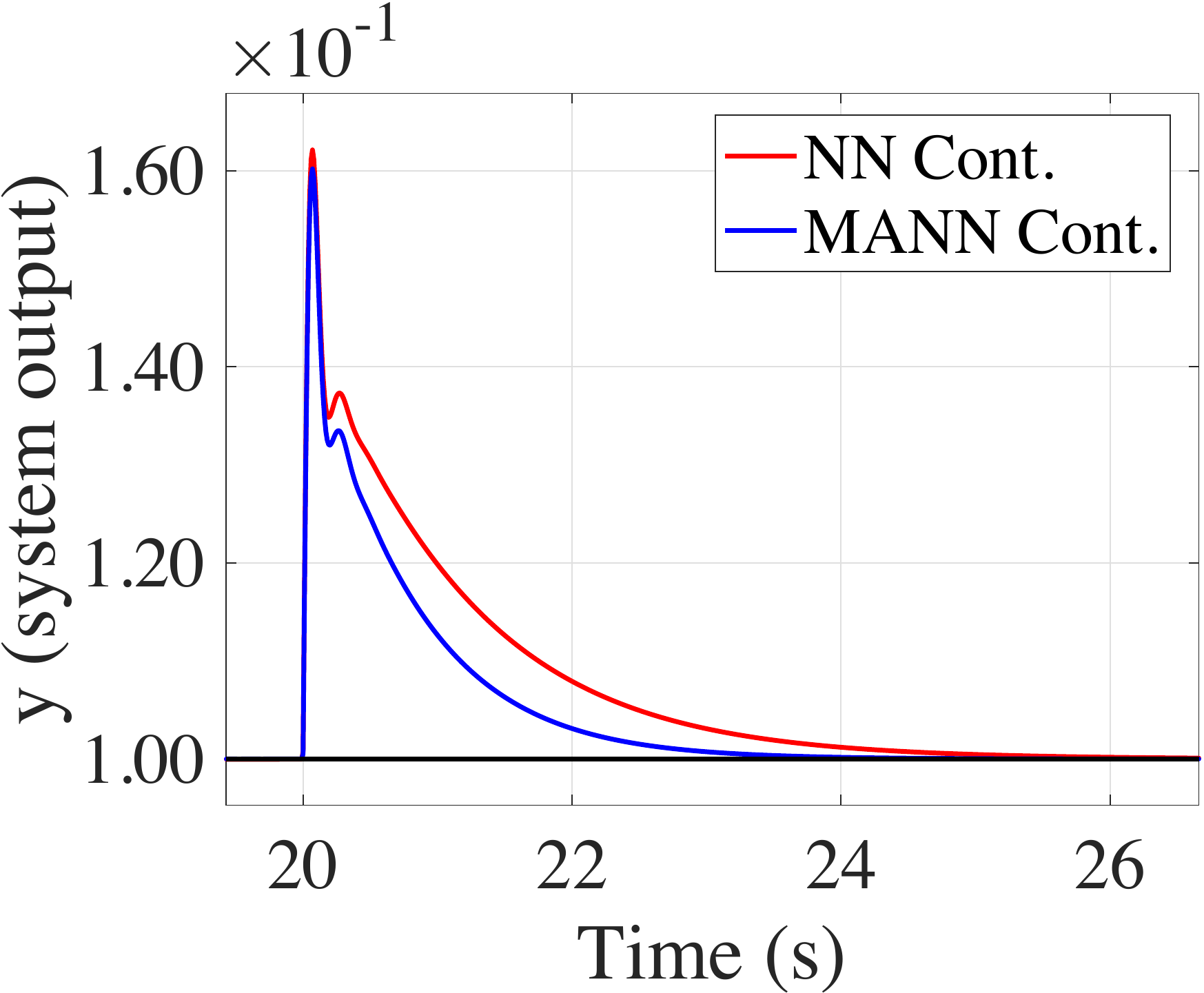} 
\end{tabular}
\caption{System response $y$ for example $1$ and scenario $3$. Top left: system response around the first two abrupt changes, top right: system response around the last abrupt change, bottom left: system response around the first two abrupt changes when $c_w = 0$, bottom right: system response around the last abrupt changes when $c_w = 0$}
\label{fig:secordsys-ex1-sc4}
\end{figure}

\begin{figure}[htp]
\begin{tabular}{ll}
\includegraphics[scale = 0.22]{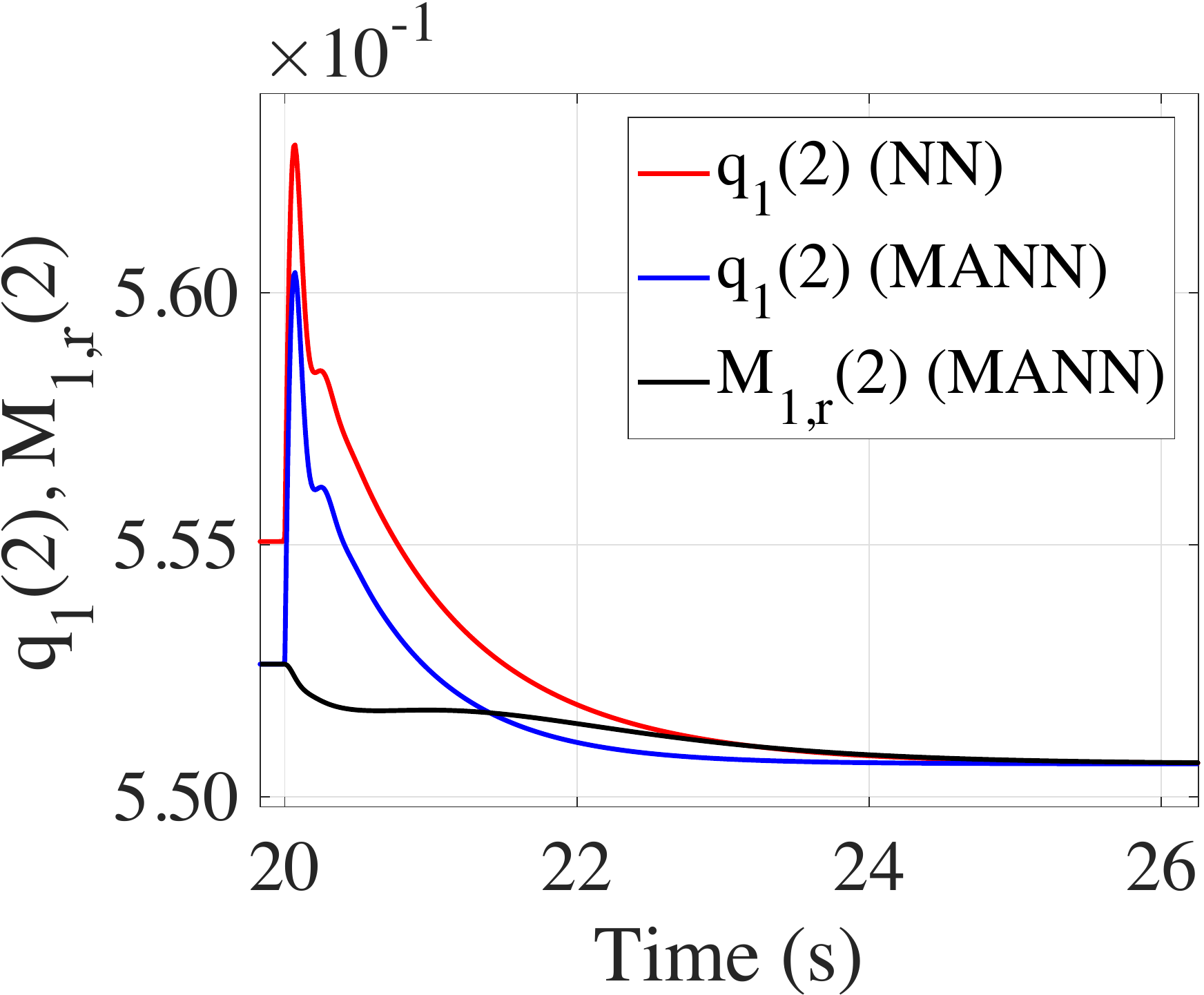} & \includegraphics[scale = 0.22]{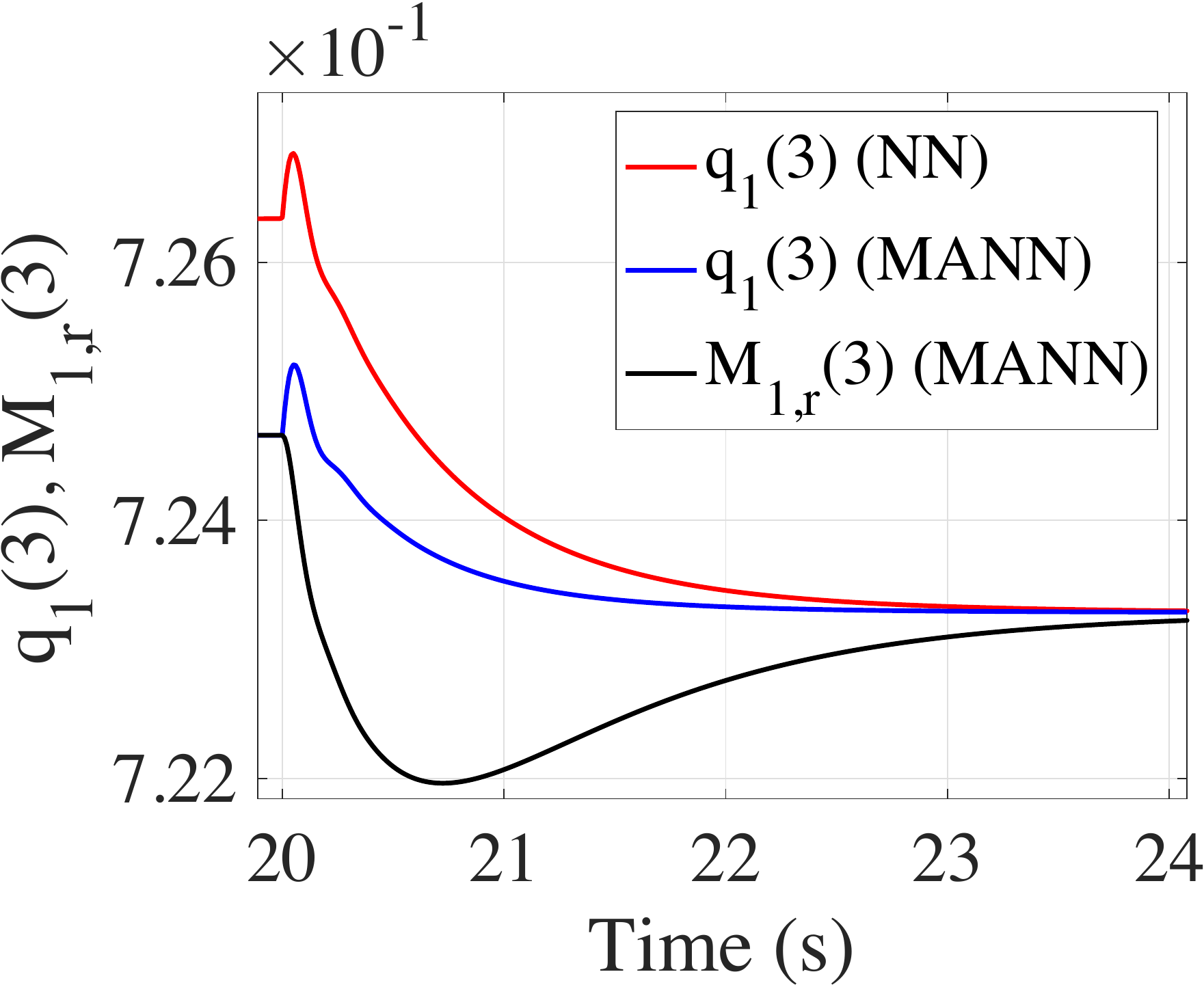}
\end{tabular}
\caption{Plots for example $1$ and scenario $3$ at third abrupt change. Top left: plot of $q_1(2)$ and $1/c_wM_{1,r}(2)$, top right: plot of $q_1(3)$ and $1/c_wM_{1,r}(3)$}
\label{fig:secordsys-ex1-sc4-error-mem}
\end{figure}

Note that the abrupt changes here are much larger than that of scenario $1$. The response plots are shown in Fig. \ref{fig:secordsys-ex1-sc4}. The plots for the NN hidden layer values for the second and third elements,  $q_1(2)$ and $q_1(3)$, and the corresponding values for memory vector scaled by $1/c_w$, i.e. $1/c_wM_{1,r}(2)$, and $1/c_wM_{1,r}(3)$ around the third abrupt change are shown in Fig. \ref{fig:secordsys-ex1-sc4-error-mem}. Note that the observations made in scenario $1$ are applicable to this scenario as well. Once again these observations are suggestive that the modification of the NN output with the memory contents that is continuously updated induces the learner of the MANN controller to converge in quick time. 

\section{Conclusion}
In this work, we proposed a backstepping memory augmented NN (MANN) adaptive control design for strict feedback nonlinear systems whose functions that determine the dynamics of the plant are completely unknown and can undergo abrupt changes. In the proposed design each NN is augmented by an external working memory. The controller can write relevant information to each of its working memory, which in this case is the hidden layer output, and retrieve them to modify its output, providing it with the capability to leverage recently learned information to improve its speed of learning. We showed through extensive simulations on multiple examples that the closed loop system that uses MANN controller recovers significantly faster after abrupt changes when compared to the regular NN controller. We also proved that the closed loop system with the MANN controller is {\it uniformly ultimately bounded}. More generally, we have demonstrated that `general principles of learning' and architectural innovations inspired from human cognition can be leveraged to improve learning in control. 

\bibliographystyle{unsrt}
\bibliography{Refs}

\section{Appendix}
\label{sec:appendix}
\subsection{Proof of Lemma \ref{lem:stabfirstorder-1}}

{\it Proof}: Differentiating $L_{e_1}$, we get,
\begin{align} 
\dot{L}_{e_1} & = e_1 \beta_1 \dot{e}_1 + \dot{y}_{d} \int_{0}^{e_1} \alpha \frac{\partial \beta_1}{\partial \alpha} d\alpha \nonumber\\
 & = e_1 \beta_1 \left(g_1(x_1)u_1 + f_1(x_1) -\dot{y}_d\right) + \dot{y}_{d} \int_{0}^{e_1} \alpha \frac{\partial \beta_1}{\partial \alpha} d\alpha \nonumber
 \end{align}

Applying UV rule for integration to the last term, we get,
\beq 
\dot{L}_{e_1}   = e_1 \left(u_1 + \beta_1 f_1(x_1) - \dot{y}_{d} \int_{0}^{1} \beta_1(\theta e_1 + y_{d}) d\theta \right)
\eeq

Using the expression for $u_1$ \eqref{eq:continpfirstorder}, we get,
\beq \dot{L}_{e_1} = -K_1e^2_1 \eeq

Then, using LaSalle's invariance principle we can conclude, for the system defined in \eqref{eq:sysfirstord} and the control input \eqref{eq:continpfirstorder}, that the closed loop system tracks the command signal asymptotically, i.e., $e_1 \rightarrow 0$ as $t \rightarrow \infty$. $\blacksquare$

\subsection{Proof of Theorem \ref{thm:stability}}

{\it Proof}: 
The derivative of $e_1$ for this case is given by,
\begin{equation}
 \dot{e}_1 = f_1(x_1) +g_1(x_1)e_2 +g_1(x_1)x_{2,d} -\dot{y}_d
\end{equation}

Consider the positive-definite function as before, i.e., $L_{e_1}$. Thus, it follows that,
\beq \dot{L}_{e_1} = -K_1 e^2_1 + e_1\left(h_1(\tilde{x}_1) - \hat{h}_1(\tilde{x}_1)\right) + e_1\mathbf{g}_1(x_1)e_2 \eeq

Define $\tilde{h}_1 = h_1 - \hat{h}_1$. Then,
\beq \dot{L}_{e_1} = -K_1 e^2_1 + e_1 \tilde{h}_1 + e_1\mathbf{g}_1(x_1)e_2 \label{eq:Le1nthorder} \eeq

Consider a second positive-definite function $L_{e_2}$, given by,
\beq L_{e_2} = L_{e_1} + \int_{0}^{e_2} \alpha \beta_2(x_1, \alpha + x_{2,d}) d\alpha \eeq

Differentiating either side w.r.t time, we get,
\begin{align} 
\dot{L}_{e_2} & = \dot{L}_{e_1} + e_2 \beta_2 \dot{e}_2 +  \dot{x}_1\int_{0}^{e_2} \alpha \frac{\partial \beta_2 (x_1, \alpha + x_{2,d})}{\partial x_1} d\alpha \nonumber \\
& + \dot{x}_{2,d} \int_{0}^{e_2} \alpha \frac{\partial \beta_2 (x_1, \alpha + x_{2,d})}{\partial \alpha} d\alpha 
\end{align}

Applying UV rule for integration to the last term, we get,
\begin{align} 
\dot{L}_{e_2} & = \dot{L}_{e_1} + e_2 \beta_2 \dot{e}_2 +  \dot{x}_1\int_{0}^{e_2} \alpha \frac{\partial \beta_2 (x_1, \alpha + x_{2,d})}{\partial x_1} d\alpha \nonumber \\
& + e_2 \beta_2 \dot{x}_{2,d} - \dot{x}_{2,d} e_2 \int_{0}^{1} \beta_2 (x_1, \theta e_2 + x_{2,d}) d\theta,
\label{eq:dVeq1nthorder} 
\end{align}


Substituting for $\dot{L}_{e_1}$ and $\dot{e}_1$ and using the expression for $h_2$ in \eqref{eq:dVeq1nthorder}, we get,
\beq
\dot{L}_{e_2} = -\sum_{i = 1}^2 K_i e^2_i + \sum_{i = 1}^2 e_i \tilde{h}_i + e_2 \mathbf{g}_2 e_3 
\label{eq:dVeq2nthorder} 
\eeq

%

Finally, consider the positive-definite function,
\beq L = L_{e_n} + \sum_{i=1}^n \left(\frac{1}{C_w} \tilde{W}^T_i\tilde{W}_i + \frac{1}{C_v} \tilde{V}^T_i\tilde{V}_i \right) \eeq

The function $L$ is a positive-definite function of $e_i$s and $\tilde{Z}_i$s. Differentiating $L$ w.r.t time and following steps similar to the proof of Theorem $3.1$ (equations A.4 to A.7) in \cite{zhang2000adaptive} we can show that,
\begin{align}
& \dot{L} \leq -\sum_{i=1}^{n} K e^2_i\left(1/2+ \int_{0}^1 \theta \mathbf{g}_i(\mathbf{x}_{i-1}, \theta e_i + x_{i,d}) d\theta \right) \nonumber\\ 
& + \sum_{i=1}^{n}-k_z \norm{e_i}^2_2 \norm{\hat{W}_i}_F\norm{\mu_i}_F -e_i\hat{W}^T_iM_{i,r} \nonumber\\
& + \sum_{i=1}^{n}-\frac{\kappa}{2}\left(\norm{\tilde{W}_i}^2_F + \norm{\tilde{V}_i}^2_F\right) + c_{i,2}, \nonumber\\
 & \text{where} \  c_{i,2} = \frac{1}{4K}\left(\norm{W^*_i}^2_2 +  \norm{V^*_i}^2_2 + \norm{W^*_i}^2_1 + \epsilon^2_i \right) \nonumber \\
 & + \frac{\kappa^2}{2} \left( \norm{W^*_i}^2_2 +  \norm{V^*_i}^2_F\right).
 \label{eq:dVeq1nthorder} 
 \end{align}
 
The difference here is that there is an additional term (the second term in the above equation). Using the fact that $e_i\hat{W}^T_iM_{i,r} \leq \norm{e_i}_2\norm{\hat{W}}_F\norm{\mu}_F$ we can rewrite above expression as:
\begin{align}
& \dot{L} \leq -\sum_{i=1}^{n} K e^2_i\left(1/2+ \int_{0}^1 \theta \mathbf{g}_i(\mathbf{x}_{i-1}, \theta e_i + x_{i,d}) d\theta \right) \nonumber\\ 
&  \sum_{i=1}^{n} - k_z \norm{e_i}^2_2 \norm{\hat{W}_i}_F\norm{\mu_i}_F + \norm{e_i}_2\norm{\hat{W}_i}_F\norm{\mu_i}_F \nonumber\\
& \sum_{i=1}^{n} -\frac{\kappa}{2}\left(\norm{\tilde{W}_i}^2_F + \norm{\tilde{V}_i}^2_F\right) + c_{i,2}
 \label{eq:dVeq2nthorder} 
 \end{align}
 
%
%
%
%


Hence, $\dot{L} < 0$ when
\beq \norm{e_i}_2 > \sqrt{\frac{2 c_{i,2}}{K(1+g_{i,0})}} \ \text{and} \ \norm{e_i}_2 > \frac{1}{k_z} \eeq

Denote $r_{i,e} = \max \left\{\sqrt{\frac{2 c_{i,2}}{K(1+g_{i,0})}}, \frac{1}{k_z}\right\}$. From the definition of the constant $c_{i,2}, k_z$ and $\kappa$ it follows that $r_{i,e} = O(1/K)$. We can simplify the first condition as follows: $\dot{L} < 0$ when $\norm{e_i}_2 >  r_{i,e}$. 

Assume for the time being that $\norm{\mu_i}_F \leq \overline{\mu}_i$. When $\norm{e_i}_2 \leq r_{i,e}$ we have that
\beq
\dot{L} \leq  \sum_{i=1}^{n} -\frac{\kappa}{2}\left(\norm{\tilde{W}_i}^2_F + \norm{\tilde{V}_i}^2_F\right) + r_{i,e}\norm{\hat{W}_i}_F\overline{\mu}_i  + c_{i,2}. \eeq

We can rewrite the above expression as 
\beq
\dot{L} \leq  \sum_{i=1}^{n} -\frac{\kappa}{2}\left(\norm{\tilde{W}_i}^2_F+ \norm{\tilde{V}_i}^2_F\right) + r_{i,e}\norm{\tilde{Z}_i}_F\overline{\mu}_i  + \tilde{c}_{i,2}, \eeq

where $\tilde{c}_{i,2} = c_{i,2} + r_{i,e}\overline{\mu}_iZ_m$. Noting that $\norm{\tilde{Z}_i}^2_F = \norm{\tilde{W}_i}^2_F+ \norm{\tilde{V}_i}^2_F$, $\dot{L} < 0$ also when
\begin{equation}
\norm{\tilde{Z}_i}_F > \sqrt{\frac{2 (c_{i,2} + r_{i,e}\overline{\mu}_iZ_m)}{\kappa}}, \text{and} \ \norm{\tilde{Z}_i}_F > \frac{2r_{i,e}\overline{\mu}_i}{\kappa}
\end{equation}
Define:
\begin{equation}
r_{i,z} = \max \left\{\sqrt{\frac{2 (c_{i,2} + r_{i,e}\overline{\mu}_iZ_m)}{\kappa}}, \frac{2r_{i,e}\overline{\mu}_i}{\kappa} \right\}
\end{equation}

Since $\kappa = \frac{1}{\sqrt{K}}$, it follows from the expression for $r_{i,e}$ that $r_{i,e} = O\left(\frac{1}{K}\right)$. This makes $r_{i,z} = O\left(\frac{1}{K^{1/4}}\right)$ and so is sufficiently small because $K$ is large. Thus, $\dot{L}$ is negative outside a compact set defined by the radii $r_{i,e}$, and $r_{i,z}$. Earlier we assumed that $\norm{\mu_i}_F$ is bounded. From the memory update equations \eqref{eq:memorywrite} it follows trivially that $\mu_i$ is bounded when $\tilde{Z}_i$ and $e_i$ are bounded. Denote this bound by $\tilde{\mu}_i$ when $\norm{e_i}_2 \leq r_{i,e}$ and $\norm{\tilde{Z}_i}_F \leq r_{i,z}$. We can set $\overline{\mu}_i \gg \tilde{\mu}_i$ to ensure consistency of the bound used in the derivation above. This establishes that $\dot{L} < 0$ outside a compact set defined by $r_{i,e}$s and $r_{i,z}$s. It follows from the UUB Lyapunov theorem (Refer \cite{lewis1998neural}) that if the signals start within these compact sets then the signals will stay within a neighborhood of this compact set. In the above steps we had assumed that the control signals are valid i.e. the NN approximation holds throughout.

Next we show that the control signals are valid provided the signals $e_i$s and $\tilde{Z}_i$s start from within the compact set defined by $r_{i,e}$s and $r_{i,z}$s. Let the compact set within which the NN approximation holds be given by radius $\tilde{r}_{i,u}$. Then, for the approximation to hold it should be that, 
\begin{align} 
& \norm{\tilde{x}_i}_2 \leq \tilde{r}_{i,u}. \nonumber\\
& \text{That is,} \ \norm{x_i}_2 + \sum_{k = 1}^{i} \norm{y^k_d}_2 + \sum_{k = 1}^{i-1} \norm{\hat{Z}_{k}}_F \leq  \tilde{r}_{i,u} \nonumber
 \end{align}

will ensure that $\norm{\tilde{x}_i}_2 \leq \tilde{r}_{i,u}$. Let, the bounds on $y^k_d$s be $y_u$, $\overline{C} = \max\{C_w,C_v\}$, $Z_m$ be an upper bound on the Frobenious norm of $Z^{*}$. Then, the weights $\hat{Z}_{k}$s and $x_i$ should be such that
\beq \norm{x_i}_2 + \sum_{k = 1}^{i-1} \norm{\hat{Z}_{k}}_F \leq \tilde{r}_{i,u} - iy_u = r_{i,u}, \eeq

to ensure that $\norm{\tilde{x}_i}_2 \leq \tilde{r}_{i,u}$. Using the fact that $\norm{\tilde{Z}_{k}}_F \geq \norm{\hat{Z}_{k}}_F - Z_m$ we can say that
\beq \norm{x_i}_2 + \sum_{k = 1}^{i-1} \norm{\tilde{Z}_{k}}_F \leq \tilde{r}_{i,u} - iy_u -(i-1)Z_m \eeq

will ensure that $\norm{\tilde{x}_i}_2 \leq \tilde{r}_{i,u}$. Assume that it has been established that  $\norm{\tilde{x}_k}_2 \leq \tilde{r}_{k,u}$ when $\norm{e_k}_2 \leq r_{k,e}$ and $\norm{\tilde{Z}_k}_F \leq r_{k,z}$ for all $k \leq i-1$. Then,
\beq \norm{x_{i,d}}_2 \leq \frac{\tilde{K}}{g_{i,0}}\overline{r}_{i-1} + \frac{\overline{g}_{i-2}}{g_{i,0}}{\overline{r}_{i-2}} + \frac{c}{g_{i,0}}(\overline{r}_{i-1}+Z_m) = r_{i,x}\eeq

where $\overline{r}_{i-1}, \overline{r}_{i-2}, \tilde{K}, \overline{g}_{i-2}$ and $c$ are constants that depend on $\tilde{r}_{k,u}, r_{k,e}, r_{k,z}$ and $\overline{\mu}_k$ for $k \leq i-1$, when $\norm{e_k}_2 \leq r_{k,e}$ and $\norm{\tilde{Z}_k}_F \leq r_{k,z}$ for all $k \leq i-1$. It is to be noted that $r_{i,x}$ is $O(1)$ when $\norm{e_k}_2 \leq r_{k,e}$ and $\norm{\tilde{Z}_k}_F \leq r_{k,z}$ for all $k \leq i-1$. Using the fact that $\norm{e_i}_2 \geq \norm{x_i}_2 - \norm{x_{i,d}}_2 \geq \norm{x_i}_2 - r_{i,x}$ we can say that
\beq \norm{e_i}_2 + \sum_{k = 1}^{i-1} \norm{\tilde{Z}_{k}}_F \leq \tilde{r}_{i,u} - iy_u -(i-1)Z_m - r_{i,x} = r_{i,u} \eeq

will ensure that $\norm{\tilde{x}_i}_2 \leq \tilde{r}_{i,u}$. Using the fact that $r_{i,u} \sim O(1)$, we can make $K$ sufficiently large such that when $\norm{e_i}_2 \leq r_{i,e} \}$ and $\norm{\tilde{Z}_i}_F \leq r_{i,z} \}$,
\beq \norm{e_i}_2 + \sum_{k = 1}^{i-1} \norm{\tilde{Z}_{k}}_F \leq r_{i,e} + \sum_{k = 1}^{i-1} r_{i,z} \ll r_{i,u} \eeq

This will ensure that $\norm{\tilde{x}_i}_2 \leq \tilde{r}_{i,u}$ when $\norm{e_i}_2 \leq r_{i,e}$ and $\norm{\tilde{Z}_i}_F \leq r_{i,z}$ given that it has been established for all $k \leq i-1$. It is trivial to establish this for $i = 1$. Hence, by the principle of induction by choosing $K$ sufficiently large we can ensure that the control signals are always valid provided the signals start from within the compact set defined by $r_{i.e}$s and $r_{i,z}$s. This completes the proof. 

\end{document}